\documentclass[twocolumn]{aastex63}

\usepackage[colorinlistoftodos]{todonotes}
\usepackage[flushleft]{threeparttable}
\usepackage{graphicx}

\let\Oldtodo\todo
\renewcommand{\todo}[1]{\Oldtodo[inline]{#1}}

\hypersetup{linkcolor=cyan,citecolor=cyan,filecolor=cyan,urlcolor=cyan}

\usepackage{amsmath}
\usepackage{float}

\shorttitle{Desert Dwellers}
\shortauthors{Hallatt \& Millholland}

\graphicspath{{./}{figures/}}

\begin{document}

\title{Shedding Light on Desert Dwellers}

\correspondingauthor{Tim Hallatt}
\email{thallatt@mit.edu}

\author[0000-0003-4992-8427]{Tim Hallatt}
\affil{MIT Kavli Institute for Astrophysics and Space Research, Massachusetts Institute of Technology, Cambridge, MA 02139, USA}

\author[0000-0003-3130-2282
]{Sarah Millholland}
\affil{MIT Kavli Institute for Astrophysics and Space Research, Massachusetts Institute of Technology, Cambridge, MA 02139, USA}
\affil{Department of Physics, Massachusetts Institute of Technology, Cambridge, MA 02139, USA}

\begin{abstract}

The ``sub-Jovian desert" ($2{\lesssim}R_{\rm p}{\lesssim}10 \ R_{\oplus}$, periods $\lesssim$3 days) is sparsely populated but no longer empty. Recent surveys have revealed that planets residing in the desert are dense (${\rho}{\gtrsim}1$ g/cm$^{3}$), massive ($\sim$10${-}$50 $M_{\oplus}$), and orbit metal-rich stars that are indistinguishable from those hosting hot Jupiters. However, their origins remain mysterious. In this work we adopt and test the hypothesis that tidal destruction of hot Jupiters can populate the sub-Jovian desert with stripped remnant planets. We first show that stars hosting desert dwellers exhibit Galactic kinematics indicative of an older population descended from those hosting hot Jupiters. We highlight that tidally-driven Roche lobe overflow (RLO) can indeed populate the desert with planets similar to those observed, but only if angular momentum transfer during RLO is inefficient (``lossy" RLO). The entire width of the sub-Jovian desert can be backfilled with the remnants of hot Jupiters that possessed their empirically inferred spread in entropy. In this picture, current desert dwellers such as LTT 9779b should be tidally decaying at an observationally testable rate of ${\sim}0.5$ ms/yr. Our theory also predicts that desert dweller host stars may rotate up to an order of magnitude more rapidly than field stars; rotation period differences may persist ${\sim}$Gyr after RLO. Lossy RLO may also manifest as a burst of IR excess that could outshine the host star for up to ${\sim}10^{3}$ yr. If these predictions are confirmed by observations, our theory indicates that desert dwellers can be leveraged to study the interiors of giant planets in exquisite detail.

\end{abstract}

\section{Introduction} \label{sec:intro}

Exoplanets exhibit a remarkable diversity of sizes, spanning gas-poor super-Earths ($1{-}1.7 \ R_{\oplus}$), and sub-Neptunes ($1.7{-}4 \ R_{\oplus}$), to gas-rich sub-Saturns ($4{-}8 \ R_{\oplus}$) and giant planets (${\gtrsim}10 \ R_{\oplus}$) \citep[e.g.][]{Howard10,howmarbry12,frefrator13,pethowmar13,petmarwin18}. Inside orbital periods ${\lesssim}$3 days however, the planet population bifurcates: planets with sizes intermediate between super-Earths and gas giants become exceedingly rare. Early analysis of \textit{Kepler} transiting planet data found this dearth of hot, intermediate-sized planets to be so pronounced so as to dub it the ``sub-Jupiter desert" \citep{szakis11}. Follow-up work confirmed that this sub-Jovian desert was indeed remarkably empty, especially in light of the favorable conditions to observe such planets in the desert if they existed \citep[e.g.][]{beanes13,mazholfai16,lunkjealb16}. Thanks to the \textit{TESS} mission however \citep[][]{ricwinvan15}, several ``desert dweller" planets have recently been discovered that lie squarely in the regime that was previously thought uninhabited \citep[e.g.][]{visbeh25,doyarmacu25}. The desert is no longer empty.

Initial characterization of the desert delineated a triangular region in mass/radius versus orbital period, opening inward from a locus at ${\sim}$5${-}$10 days; these boundaries separated gas giants above from super-Earths below by a region virtually devoid of planets \citep[][]{mazholfai16}. Theoretical work suggests that the desert's bottom boundary can be carved out by photoevaporative mass loss catalyzed by stellar high energy emission \citep[e.g.][]{owejac12}, as low/intermediate mass planets readily lose their atmospheres at such tight orbital distances to become denuded rocky cores comprising the lower desert edge \citep[][]{owelai18,hallee22}. The desert's upper boundary however is unlikely to be sculpted by evaporation as massive planets resist significant mass loss \citep[][though see \cite{kurnak14} for a differing view, and \cite{tholeelop23} for consideration of exceptionally low density Saturns which may evaporate in the presence of extra internal heating]{owelai18,ionpav18,visknugre22}. Instead, \cite{owelai18} point to high eccentricity migration as a means to sculpt the desert's upper edge; in this picture, the boundary reflects the tidal disruption limit where planets may safely park after their orbits circularize. Alternatively, both upper and lower desert boundaries may be set by the tidal disruption barrier which scales differently with planet mass for large and small planets, respectively \citep[][]{matkon16}. In-situ formation of hot Jupiters could also help to create the desert's upper edge \citep[][]{baibat18}.

Recent work has brought the desert and its denizens into sharper focus. Bias-correcting the desert boundaries places the outer locus at ${\sim}$3.2 days \citep[][roughly corresponding to an irradiation flux ${\sim}550{\times}$ that of the Earth; \cite{magcovcor24}]{casboulil24} and suggests differing formation pathways between bona fide desert dwellers (${<}3.2$ days) and those in the neighboring regime beyond the desert edge \cite[the ``ridge" and ``savannah";][]{bouattmall23}. In particular, \cite{visbeh25} showed that desert dweller host stars are preferentially metal-rich and have a metallicity distribution that is indistinguishable from that of hot Jupiter host stars \citep[see also][who first pointed out a preference for metal-rich host stars among hot Neptunes]{dongxiezho18}. Desert dwellers might also share the same elevated stellar multiplicity rates as hot Jupiters, unlike longer period planets \citep[][]{eelarm25}. \cite{visbeh25} highlight that shared host stellar properties between desert dwellers and hot Jupiters is consistent with a ``top-down" picture in which desert dwellers are the descendants of hot Jupiters (albeit with a possible difference in host star ages; see Section \ref{subsec:kinematics}). In this picture, hot Jupiters could be destroyed either during their high eccentricity migration or over longer timescales due to tidal inspiral. \cite{doyarmacu25} furthermore find that desert dwellers boast high densities consistent with the absence of gaseous atmospheres, in stark contrast with planets beyond the desert edge that frequently host sizable gas envelopes. Massive, dense planets devoid of gas are also suggestive of the exposed cores of giant planets \citep[][]{armlopadi20}. 

In this paper, we explore whether tidal destruction of hot Jupiters can populate the sub-Jovian desert with gas-poor remnant planets similar to those observed. We focus on the tidal inspiral channel, rather than destruction during high eccentricity migration. The tidal inspiral picture is supported by at least three empirical results. First, it is empirically known that hot Jupiters orbit stars that are kinematically young, strong evidence that they do indeed decay to destruction while their host stars are on the main sequence \citep[][see also \cite{miymas23}]{hamsch19}. Second, Sun-like stars hosting hot Jupiters are observed to spin more rapidly than those with more distant giant planets or lower mass planets \citep[][]{tejwinand21}, as expected under tidal orbital decay. Third, tidal decay has been directly measured in the WASP 12 b system \citep[][]{yeewinknu20}, which exhibits spectroscopic signatures of possible mass loss induced by Roche lobe overflow (RLO; \cite{foshasfro10,laihelheu10,antgoo22}). Since hot Jupiters are likely susceptible to tidal inspiral, we simply ask: can the remnant planets that emerge after tidal decay and subsequent RLO backfill the desert? We will show that the entire sub-Jovian desert out to ${\sim}$3 days can indeed be populated by the remnants of hot Jupiters.

This paper is structured as follows. In Section \ref{subsec:kinematics} we show that the kinematics of desert dweller host stars may indicate an older age than those hosting hot Jupiters. This result suggests an evolutionary link between hot Jupiter and desert dweller hosts, further motivating exploration of the top-down picture for desert dweller formation. Section \ref{sec:methods} outlines our theoretical calculations, while Section \ref{sec:results} highlights our results. Observational signatures of the mechanism outlined in this paper are discussed in Section \ref{sec:discussion}, and we itemize pressing issues future work should address in Section \ref{subsec:future_work}. The conclusions of this paper are summarized in Section \ref{sec:conclusion}.

\section{Galactic Kinematics of Desert Dweller Host Stars}\label{subsec:kinematics}

\begin{figure}
\epsscale{1.2}
\plotone{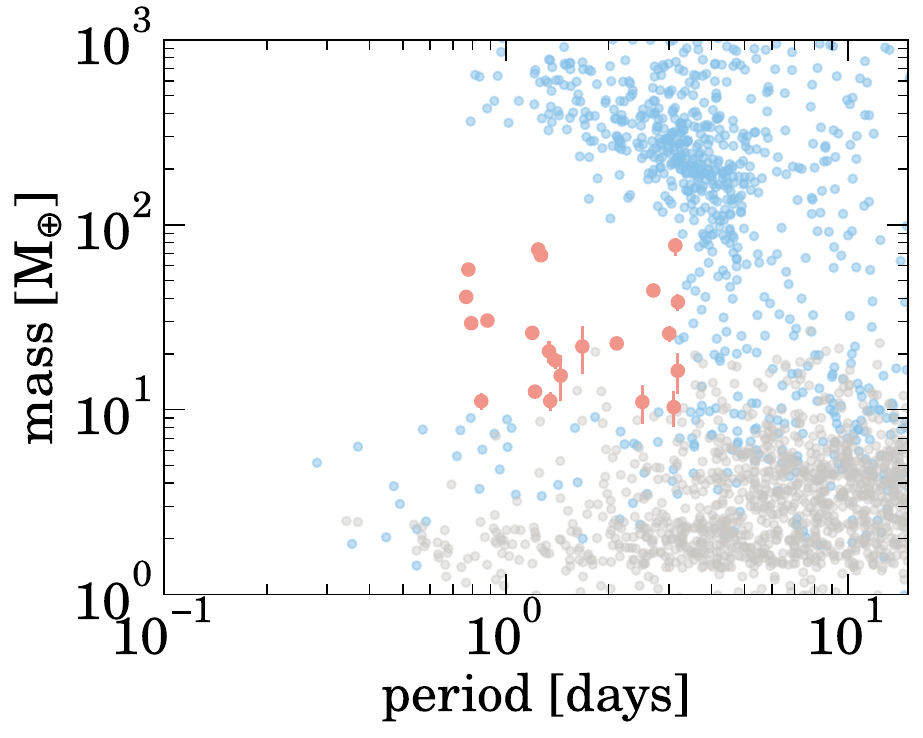}
\caption{Exoplanetary mass versus orbital period. Planets in the sub-Jovian desert (${\lesssim}3.2$ days, $10{\leq}m_{\rm p}{\leq}10^{2} \ M_{\oplus}$) are highlighted in red, while all other planets with mass measurements are in blue. Grey points display estimated masses for planets with only radius measurements following the mass/radius relation of \cite{chekip17} (planets that fall in the sub-Jovian desert after this procedure are not included in our sample). Data are from \href{https://www.astro.keele.ac.uk/jkt/tepcat/}{tepcat} \citep[][]{sou11}; some desert dwellers' data have also been augmented from the \href{https://exoplanetarchive.ipac.caltech.edu}{Exoplanet Archive}. \label{figure:desert}}
\end{figure}

If desert dwellers are the descendants of hot Jupiters, we would expect their host stars to be systematically older than those hosting hot Jupiters. As demonstrated by \cite{hamsch19}, hot Jupiter host stars exhibit a significantly lower velocity dispersion compared to that of field stars; since the stellar velocity dispersion grows with age due to dynamical heating in the Galactic disk \citep[e.g.][]{bintre08,holnorand09}, this low velocity dispersion indicates that stars hosting hot Jupiters are younger than field stars. \cite{hamsch19} propose that stars hosting hot Jupiters are found to be preferentially young due to tidal destruction of hot Jupiters while their host stars are on the main sequence. In this picture, only systems young enough to have avoided tidal inspiral can host hot Jupiters. An older stellar population should therefore exhibit a velocity dispersion in closer agreement to that of field stars than the $4\sigma$ discrepancy between hot Jupiter hosts and field stars reported by \cite{hamsch19}. The goal of this section is to demonstrate that desert dweller host stars do indeed exhibit a velocity dispersion consistent with that of field stars, indicative of an older age compared to hot Jupiter hosts. 

We first defined our sample of desert dwellers following a similar procedure as \cite{visbeh25}. The sample includes confirmed planets from the \href{https://www.astro.keele.ac.uk/jkt/tepcat/}{tepcat} catalog \citep[][downloaded 11/22/2024]{sou11} with masses satisfying $10{\leq}m_{\rm p}{<}10^{2}  \ M_{\oplus}$ and periods ${<}$3.2 days. We adopt this definition of the desert rather than previous less restrictive definitions \citep[e.g.][]{mazholfai16} to reflect updated desert boundaries based on where planet frequency appears to flatten \citep{casboulil24}, as well as to remain consistent with the population found by \cite{visbeh25} to orbit stars with similar metallicities to hot Jupiter hosts. Planets orbiting stars less massive than $0.7 \ M_{\odot}$ were excluded so that our sample only contained FGK stars. Several planets' parameters listed in \textit{tepcat} were manually updated with new, refined measurements from the  \href{https://exoplanetarchive.ipac.caltech.edu}{Exoplanet Archive}. We also added the planet TOI-1410.01 \citep[][]{pollubbea24} to the sample since it was not recorded in \textit{tepcat} despite its desert dweller status. Figure \ref{figure:desert} displays our desert dweller sample of 22 planets in context with the rest of the close-in exoplanet population. 

We next constructed a field star sample to be kinematically compared against our desert dwellers. A sample of $10^{7}$ stars with six dimensional astrometric solutions that satisfied the quality cuts outlined in the appendix of \cite{hamsch19} were downloaded from Gaia Data Release 3 \citep{gaibroval21}. We rejected evolved stars by only accepting stars that lie less than one absolute G band magnitude above the Pleiades main sequence shown in Figure 1 of \cite{hamsch19}. Our calculations did not account for extinction and reddening. The method outlined in \cite{halwie20} was used to convert the stellar astrometry to Galactic UVW velocities via \texttt{Astropy} \citep[][]{astropy22} and to compute the Galactic oscillation amplitude $z_{\rm max}$ of each star above/below the Milky Way midplane. We also computed $z_{\rm max}$ and UVW velocities for desert dweller host stars.

To compare the velocity dispersion of desert dweller host stars to that of field stars, we constructed $10^{4}$ Monte Carlo ensembles of field stars that share the same $z_{\rm max}$ and color (BP-RP Gaia bands; 330${-}$680 and 640${-}$1050 nm respectively) distributions as that of the desert dweller host stars, using a similar procedure as \cite{hamsch19}. Matching $z_{\rm max}$ helps control for hot Jupiter occurrence correlations with age \citep[e.g.][]{chexiezho23}, metallicity \citep[e.g.][]{fisval05}, and possible thin/thick disk membership \citep[e.g.][]{swabannar23,hallee25}.\footnote{We experimented with matching against metallicity instead of $z_{\rm max}$ (using \texttt{mh$\_$gspphot} in Gaia DR3) and find the same qualitative conclusion. While metallicities from Gaia DR3 are known to be biased \citep{andfousor23}, we do not expect such a bias to affect the conclusion of this test.} To match $z_{\rm max}$, we first randomly drew $10^{3}$ $z_{\rm max}$ values from the empirical (cumulative) $z_{\rm max}$ distribution function of desert dweller hosts using an inverse transform sampling technique. For each $z_{\rm max}$ we then selected the field star with the closest matching oscillation amplitude. To match the color distribution, we then selected from our $z_{\rm max}$-matched sample the 22 stars with colors closest to that of each of the 22 desert dweller host stars. The velocity dispersion for each ensemble was computed via $\sigma_{UVW}{=}N^{-1}_{\star}\Sigma_{i}[(U_{i}{-}{\langle}U{\rangle})^{2}+(V_{i}{-}{\langle}V{\rangle})^{2}+(W_{i}{-}{\langle}W{\rangle})^{2}]^{1/2}$, where $N_{\star}{=}22$ is the number of stars in our desert dweller sample (${=}$the number of stars in each Monte Carlo ensemble) and $\langle \ldots \rangle$ denotes mean quantities.

\begin{figure}
\epsscale{1.1}
\plotone{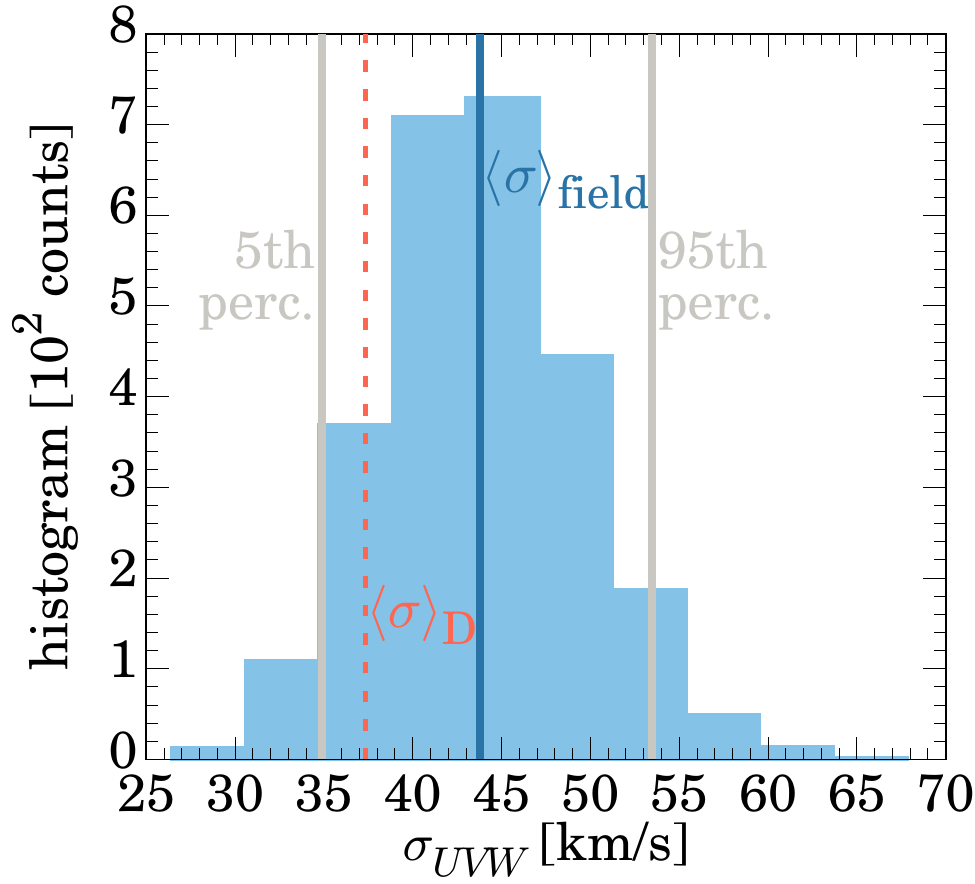}
\caption{Galactic velocity dispersion of desert dweller host stars $\langle\sigma\rangle_{\rm D}$ (dashed red vertical line), the velocity dispersion distribution from ensembles of field stars matched to desert dweller hosts (blue histogram; 5th and 95th percentile confidence intervals are shown in grey), and the mean velocity dispersion of field stars $\langle\sigma\rangle_{\rm field}$ (solid blue vertical line). The \cite{hamsch19} field star distribution has similar mean to ours but much smaller variance, which allows hot Jupiter hosts to be clearly distinguished from field stars. Our large field star variance precludes such easy distinction between field stars and desert dweller hosts; unlike hot Jupiter host stars, the velocity dispersion of desert dweller hosts is statistically indistinguishable from that of field stars. The Galactic kinematics of desert dweller host stars are consistent with an older population than those hosting hot Jupiters. \label{figure:uvw}}
\end{figure}

The velocity dispersion of desert dweller host stars is compared to that of our $z_{\rm max}$ and color-matched field star ensembles in Figure \ref{figure:uvw}. The spread in $\sigma_{UVW}$ computed from our field star ensembles is much larger than that reported by \cite{hamsch19} owing to the small number statistics of our desert dweller sample ($N_{\star}{=}22$ vs. $N_{\star}{=}338$, yielding a standard deviation ${\propto}1/\sqrt{N}{\sim}4{\times}$ larger). The velocity dispersion of desert dweller host stars is lower than that of field stars but only at a ${<}2{\sigma}$ level. Desert dweller host stars are therefore statistically indistinguishable from field stars at the current sample size. Given that desert dweller hosts share the same metallicity distribution as hot Jupiter hosts \citep{visbeh25}, their kinematics are consistent with an older population descended from those hosting hot Jupiters. The remainder of this paper serves to determine whether desert dwellers can indeed emerge from the destruction of hot Jupiters.

\section{Methods}\label{sec:methods}

The evolution of tidally disrupting gas giants depends simultaneously on planetary dynamical and structural evolution. We employ the method outlined in Paper I to couple interior structure/thermal evolution with orbital dynamical evolution. We first briefly summarize the structure/thermal evolution model in Section \ref{subsec:summary_paper1}. Readers interested in the microphysics employed in the models or benchmarking its performance against other codes should consult Paper I. Section \ref{subsec:massloss} details how we additionally account for mass loss from the planet, while Section \ref{subsection:dynamics} describes our treatment of RLO orbital dynamics.

\subsection{Internal Structure \& Thermal Evolution}\label{subsec:summary_paper1}

Following Paper I we employ the ``stepping through the adiabats" approach to follow planetary thermal and structural evolution \citep[][]{hub77,forhub03,arrbil06}. This approach requires that the convective turnover time (${\sim}$weeks) be shorter than the global cooling and/or mass loss timescale of the planet so that the convective zone remains adiabatic as the planet evolves \citep[e.g.][]{pacsie72,arrbil06}. This approach also requires that the thermal relaxation time in the radiative zone, $t_{\rm rad}{\sim}c_{\rm P}T\rho H/F{\sim} c_{\rm P} TP/g F{\sim}10^{3-4}$ yr (with $c_{\rm P}$ the specific heat capacity, $H{=}c^{2}_{\rm s}/g$ the scale height, $c_{\rm s}$ the sound speed, $g$ the gravitational acceleration, and $F$ the local flux; see Paper I for detailed calculation of $t_{\rm rad}$), is shorter than the global cooling/mass loss time so that energy flows outward at constant rate $L$ through the radiative zone as the planet evolves. Our calculations generally satisfy both of these assumptions.\footnote{We do find these assumptions to be violated in the most extreme cases we consider when the mass loss time ${\sim}m_{\rm p}/|\dot{m}_{\rm p}|$ at the peak of Roche lobe overflow falls shorter than the radiative time. We have verified that capping the mass loss rate so that the radiative and convective timescales are always less than the mass loss time, i.e. ensuring the planet always has ample time to adjust to the mass loss, does not change our results.}

To follow the evolution, a large grid of planetary structure models in thermal and mechanical equilibrium were pre-computed in Paper I. For every planet model, we tabulated the specific entropy $S$ of the innermost convective zone, the integral of $\int_{\rm conv}Tdm$ across the convective zone (with $T$ the temperature and $dm$ the mass increment; see equation \ref{equation:cooling}), the luminosity $L$ (which is spatially constant under our assumptions), and the photospheric radius $R_{\rm P}$ (defined where the optical depth to internal radiation is 2/3). Under our assumptions listed above, the entropy equation $\partial L/\partial m=-T\partial S/\partial t$ reads \citep[e.g.][]{marcum14},

\begin{equation}\label{equation:cooling}
    \frac{dS}{dt}=\frac{-L}{\int_{\rm conv} T dm}.
\end{equation}

\noindent We integrate equation \ref{equation:cooling} by looking up $L$ and $\int_{\rm conv}Tdm$ from our tables on the fly as the planet cools. 

We have not included in equation  \ref{equation:cooling} a cooling term due to advection of thermal energy from the convective zone during mass loss, which would decrease $S$ \citep[][]{owewu16,tanformur24}. As shown by \cite{tanformur24}, since convection transports energy and material much in excess of that required to feed the wind (see Paper I for estimates of convective turnover times, which are indeed $\ll$mass loss times throughout this work), the convective zone readjusts to the mass loss adiabatically. The outflow decreases the total energy budget of the convective zone by advecting away internal and gravitational binding energy, without changing the thermal state/entropy of material remaining in the convective zone.

Our structure grid has five axes along which planets may vary. First we use two core masses $m_{\rm c}{\in}\{10,20\} \ M_{\oplus}$, with radii set by the mass-radius curves of \cite{formarbar07} for Earth-like composition. Second we vary planet mass from $m_{\rm p}{\in}[10.1{-}300] \ M_{\oplus}$. The grid uses steps of $\Delta m_{\rm p}{=}0.1 \ M_{\oplus}$ for atmospheres ${\leq}1 \ M_{\oplus}$, and steps of $\Delta m_{\rm p}{=}2 \ M_{\oplus}$ otherwise (a balance between speed of computing/interpolating the grid without loss of accuracy). Third we explore two atmospheric metal mass fractions $Z$: solar $Z{=}0.02$ (with hydrogen mass fraction $X{=}0.74$) and super-solar $Z{=}0.5$ ($X{=}0.36$). Fourth we vary the entropy $S$, ranging from ${\sim}6{-}11 \ k_{\rm B}/m_{\rm H}$ for $Z{=}0.02$, and ${\sim}4{-}8 k_{\rm B}/m_{\rm H}$ for $Z{=}0.5$ ($k_{\rm B}$ is Boltzmann's constant and $m_{\rm H}$ is the hydrogen mass). Lastly we vary the equilibrium temperature at the outer boundary, which partly sets the cooling rate by controlling the location of the radiative convective boundary \citep[e.g.][]{arrbil06}; we choose three discrete equilibrium temperatures $T_{\rm eq}{\in}\{2884,2039,1442\}$ K corresponding to distances 0.01, 0.02, and 0.04 au from a Sun-like star respectively (also chosen to balance speed of computing and accuracy of interpolation). These $T_{\rm eq}$ were chosen to bracket the range of orbital distances traversed by our hot Jupiters/hot Neptunes during tidal disruption, so that we interpolate within this temperature range throughout our calculations. The planet models use the SCvH hydrogen/helium equation of state \citep[][]{saucha95}, a high-metallicity equation of state for SiO$_{2}$ from \texttt{QEOS} \citep[][]{faitauios18} and opacities from \cite{freelusfor14}.

\subsection{Mass Loss}\label{subsec:massloss}

We next detail how we account for hydrodynamic atmospheric escape. We track two flavors of hydrodynamic mass loss: Roche lobe overflow (RLO), catalyzed by bolometric heating from the central star, and photoevaporation (PE), driven by ``soft" X-ray ($0.1{\lesssim}h\nu{\lesssim}$1-2 keV) and EUV ($h\nu{\gtrsim}13.6$ eV) stellar photons (``XUV" for X-ray plus EUV). Since RLO and PE influence the evolution of close-in planets in different ways (which we outline below), it is critical to correctly identify when RLO transitions to PE and vice versa. In Sections \ref{subsubsection:rlo} and \ref{subsubsection:pe}, we first recap the physics underpinning each mass loss channel, and then in Section \ref{subsubsection:rlovpe} we establish under what conditions planets are likely to transition between RLO and PE.

\subsubsection{Roche lobe overflow}\label{subsubsection:rlo}

When the planet radius is commensurate with the Roche surface, atmospheric gas flows from the photosphere to the first Lagrange point (L1) before ultimately being transferred to the star or lost from the system \citep[e.g.][]{priwad85}. In the standard picture, mass transfer is mediated by an accretion disk which can tidally back-react with the planet \citep[e.g.][]{ver82,jacjenpea16}. We review the dynamical effect of mass transfer in Section \ref{subsection:dynamics}.

When the planetary photosphere underfills the Roche surface, the isothermal, bolometrically-heated gas escapes as a transonic wind with a sonic point at L1 \citep[e.g.][]{lubshu75,rit88}. The wind passes L1 through a nozzle that is hydrostatic perpendicular to the flow direction onto the star, with cross sectional area ${\sim}(c_{\rm s}/\Omega)^{2}$ \citep{lubshu75} where $\Omega$ is the orbital frequency and $c_{\rm s}{=}(k_{\rm B}T/\mu m_{\rm H})^{1/2}$ is the sound speed with $\mu$ the mean molecular weight and $T$ the photospheric temperature. The mass loss rate is, 

\begin{equation}\label{equation:mdot_rlo}
    \dot{m}_{\mathrm{RLO}}=-\frac{2\pi}{\sqrt{\mathcal{A}(\mathcal{A}-1)}}c_{\rm s}\rho_{\rm L1}\bigg(\frac{c_{\rm s}}{\Omega}\bigg)^{2}; \ R_{\rm p}{\leq}R_{\rm R},
\end{equation}

\noindent where the prefactor corrects for the elliptical geometry of the nozzle at L1 \citep{jacarrpen17}, and the geometric factor $\mathcal{A}$ depends on the mass ratio $q{=}m_{\rm p}/M_{\star}$ via \citep{jacarrpen17},

\begin{equation}\label{equation:A}
    \mathcal{A}=4+\frac{4.16}{-0.96+q^{1/3}+q^{-1/3}}.
\end{equation}

\noindent The density at L1, $\rho_{\rm L1}$, can be estimated using the flow's Bernoulli integral \citep{jed69,pacsie72}. Since the gas is launched strongly subsonically, this yields

\begin{equation}\label{equation:rho_roche}
    \rho_{\rm L1}=\frac{\rho_{\rm ph}}{\sqrt{e}}\exp{\bigg(-\frac{\phi_{\rm L1}-\phi_{\rm ph}}{c^{2}_{\rm s}}
    \bigg)},
\end{equation}

\noindent where $\rho_{\rm ph}$ is the density at the planet's photosphere and $\phi_{\rm L1}-\phi_{\rm ph}$ is the potential difference between L1 and photosphere.\footnote{We have explicitly verified that including the dissociation of H$_{2}$ in the Bernoulli integral changes RLO mass loss rates by order unity factors and therefore does not alter our results significantly. See Section \ref{sec:appendix_bernoulli} for the details of our calculation.} We estimate the potential difference using the expansion from \cite{jacarrpen17}, which applies to any mass ratio $q$:

\begin{align}\label{equation:jac_potential}
\begin{split}
\phi(r)&=-\frac{GM_{\star}}{a}\bigg(1+\frac{1}{2(1+q)}\bigg)\\
       &-\frac{Gm_{\rm p}}{r}\bigg[1+\frac{1}{3}\frac{1+q}{q}\bigg(\frac{r}{a}\bigg)^{3}
       \\
       &+\frac{4}{45}\bigg(1+\frac{5}{q}+\frac{13}{q^{2}}\bigg)\bigg(\frac{r}{a}\bigg)^{6}\bigg].
\end{split}
\end{align}

\noindent where $r$ is the radial coordinate of the surface at which to evaluate the potential and $a$ is the planet's semimajor axis. Since equipotential surfaces in the Roche problem are not spherical, the potential at the Roche surface $\phi_{\rm L1}$ is estimated at the volume-equivalent Roche radius $R_{\rm R}$, defined as the radius of a sphere with volume equal to that enclosed by the Roche surface \citep[][]{egg83}:

\begin{equation}\label{equation:roche_radius}
    R_{\rm R}=a \frac{0.49q^{2/3}}{0.6 q^{2/3}+\ln(1+q^{1/3})},
\end{equation}

\noindent The potential difference is then $\phi_{\rm L1}{-}\phi_{\rm ph}{=}\phi(R_{\rm R}){-}\phi(R_{\rm p})$, since the volume-equivalent radius of the photosphere lies within ${\sim}1\%$ of the photospheric radius $R_{\rm p}$ \citep{jacarrpen17}.

Equation \ref{equation:mdot_rlo} applies to planets that underfill the Roche lobe ($R_{\rm p}{<}R_{\rm R}$). In the case that the planet's photosphere exceeds the Roche lobe, we assume that the portion of the atmosphere above the Roche surface flows adiabatically along equipotential streamlines to L1. This adiabatic assumption holds if the optically thick gas below the photosphere remains optically thick to incoming radiation as it flows toward L1, and in the absence of internal dissipation. The mass loss rate reads \citep[][]{kolrit90,cehpuj23},

\begin{align}
\begin{split}\label{equation:mdot_rlo_overflow}
    \dot{m}_{\mathrm{RLO}}&=\dot{m}_{\mathrm{RLO}}\big|_{R_{\rm p}=R_{\rm R}}\\
    &\hspace*{-1cm}+\frac{2\pi}{\sqrt{\mathcal{A}(\mathcal{A}-1)}}\frac{1}{\Omega^{2}}\int_{P_{\rm ph}}^{P_{R}}G(\Pi)\bigg(\frac{k_{\rm B}T(P)}{\mu(P) m_{\rm H}}\bigg)^{1/2}dP; \ R_{\rm p}{>}R_{\rm R},
\end{split}
\end{align}

\noindent where the integral spans from the photospheric pressure down to the pressure at $R_{\rm R}$, and $G(\Pi){=}\Pi^{1/2}(\frac{2}{\Pi+1})^{\frac{\Pi+1}{2(\Pi-1)}}$ is a function of the gas' adiabatic exponent ${\Pi{=}\partial \log P/\partial \log \rho |_{S}}$ (tabulated in our EOS from Paper I). The first term in equation \ref{equation:mdot_rlo_overflow} accounts for saturated optically thin mass loss for which we evaluate equation \ref{equation:mdot_rlo} with $\rho_{\rm L1}{=}\rho_{\rm ph}$. The second term tracks the adiabatic loss of optically thick material above $R_{\rm R}$ but below $R_{\rm p}$; equation \ref{equation:mdot_rlo_overflow} becomes equation \ref{equation:mdot_rlo} when $R_{\rm p}{=}R_{\rm R}$. We calculate the second term in equation \ref{equation:mdot_rlo_overflow} by integrating the stellar structure equations (equations 1 in Paper I) inward from our photospheric boundary conditions tabulated in our planet structure grid to obtain $\Pi(P), \ T(P)$ and $\mu(P)$. The integration is performed using scipy's \texttt{BDF} solver to default precision \citep[][]{virgomoli20}.

\subsubsection{Photoevaporation}\label{subsubsection:pe}

X-ray and EUV photons from the central star can also drive hydrodynamic escape by liberating electrons in planetary upper atmospheres that heat the gas to near escape speed. Adopting for simplicity the ``energy-limited" approximation, the mass loss rate reads,

\begin{equation}\label{equation:mdot_pe}
    \dot{m}_{\rm PE}=-\eta\frac{L_{\rm PE}R^{3}_{\rm p}}{4 a^{2}G m_{\rm p}\mathcal{K}_{\rm t}}\bigg(\frac{Z}{Z_{\odot}}\bigg)^{-0.77},
\end{equation}

\noindent with $\eta{=}0.17$ the mass loss ``efficiency" \citep[e.g.][]{sheionlam14,wu19} and $L_{\rm PE}$ the stellar luminosity of high energy photons that drive PE (X-ray $L_{\rm X}$ or XUV $L_{\rm XUV}$). The $\mathcal{K}_{\rm t}$ term corrects for tidal enhancement of the flow when the planet radius approaches the Roche lobe ($\mathcal{K}_{\rm t}{\rightarrow}1$ when $R_{\rm p}{\ll}R_{\rm R}$, and $\mathcal{K}_{\rm t}{\rightarrow}0$ when $R_{\rm p}{\rightarrow}R_{\rm R}$). \cite{erkkullam07} derived $\mathcal{K}_{\rm t}$ by writing the potential difference between the planet and its Roche lobe as an expansion in the distance between L1 (located at the Hill radius $R_{\rm H}{=}a(m_{\rm p}/3 M_{\star})^{1/3}$) and the planet radius. \cite{jacarrpen17} however show that a more accurate approach that corrects for the non-spherical Roche geometry is to compute the potential difference following equations \ref{equation:jac_potential} and \ref{equation:roche_radius} (and surrounding discussion). We therefore adjust the expression for the potential difference derived by \cite{erkkullam07} ($\Delta \Phi$ in their equation 12) by taking 

\begin{equation}\label{equation:ktide}
    \mathcal{K}_{\rm t}=\frac{|\phi_{\rm L1}-\phi_{\rm ph}|}{Gm_{\rm p}/R_{\rm p}}
\end{equation}

\noindent This approach is consistent with the fact that when $R_{\rm p}{\sim}R_{\rm R}{<}R_{\rm H}$, the evaporation rate will diverge since the planet photosphere lies on the same equipotential as the Roche lobe. The metallicity scaling we adopt derives from X-ray photoevaporation of protoplanetary disks for which metal line cooling (primarily carbon and oxygen) damps the evaporation rate \citep{erccla10}. Higher $Z$ should also decrease the mass loss rate for EUV-driven flows, since the heating rate is set by hydrogen photoionization while cooling is still dominated by metals \citep{owemur18}. Dedicated studies determining the exact scaling for $\dot{m}_{\rm PE}$ with $Z$ in the context of planetary evaporation would be helpful to improve our formulation.

We now outline how we determine whether X-ray or EUV photons drive the mass loss. Evaporation is driven by EUV photons when the EUV ionization front lies close enough to the planet that the X-ray heated gas below cannot pass through a sonic point \citep[the X-ray absorption cross section is ${\sim}10^{4}{\times}$ smaller than that of EUV photons, so X-rays penetrate much deeper;][]{owejac12}. To decide between X-ray or EUV we therefore must check whether an X-ray driven outflow is optically thick to EUV photons down to the X-ray sonic surface. The optical depth to EUV photons from L1 down to the X-ray wind's sonic point is

\begin{equation}\label{equation:tau_euv}
    \tau_{\rm EUV}=\int^{r_{\rm s}}_{\rm L1}n_{\rm H}(r)\sigma_{\rm EUV}dr,
\end{equation}

\noindent where $r_{\rm s}$ is the radius of the X-ray sonic surface, $n_{\rm H}$ is the number density of hydrogen in the outflow, and $\sigma_{\rm EUV}{=}1.9{\times}10^{-18}$ cm$^{2}$ is the EUV photoionization cross section evaluated for a typical EUV energy of 20 eV \citep[][]{spi78}. The hydrogen number density in the X-ray wind follows from mass continuity; assuming the X-ray heated gas outflows at a rate given by equation \ref{equation:mdot_pe}, this yields

\begin{equation}\label{equation:continuity}
n_{\rm H}(r)=\frac{\dot{m}_{\rm PE,X}}{4\pi r^{2} c_{\rm s, X}\mathcal{M}_{\rm X}(r)}\frac{X}{m_{\rm H}},
\end{equation}

\noindent where we reiterate for clarity that $X$ is the hydrogen mass fraction of the gas, $c_{\rm s,X}=(k_{\rm B}T_{\rm X}/\mu_{\rm X}m_{\rm H})^{1/2}{\sim}6$ km s$^{-1}$ is the sound speed for X-ray heated gas at temperature $T_{\rm X}{\sim}5{\times}10^{3}$ K \citep[e.g.][]{glanajige04}, and $\mathcal{M}_{\rm X}$ is the X-ray flow's Mach number. To evaluate $\dot{m}_{\rm PE, \rm X}$ we set $L_{\rm PE}{=}L_{\rm X}$. We estimate $r_{\rm s}$ and $\mathcal{M}_{\rm X}$ assuming the X-ray heated gas is isothermal at temperature $T_{\rm X}$. Following the procedure outlined by \cite{par58} \cite[see also][]{lamcas99}, combining mass continuity and momentum balance yields the following equation for the flow speed $u$:

\begin{equation}\label{equation:parker_tidal}
    \left(u-\frac{c^{2}_{\rm s,\rm X}}{u}\right)\frac{\partial u}{\partial r}=2\frac{c^{2}_{\rm s,\rm X}}{r}-\frac{Gm_{\rm p}}{r^{2}}+3\frac{GM_{\star}r}{a^{3}}.
\end{equation}

\noindent The last term in equation \ref{equation:parker_tidal} accounts for centrifugal and differential stellar gravity acceleration in the frame rotating with the planet's orbital frequency. At the sonic point $u{=}c_{\rm s,X}$ so that the X-ray sonic point is given by the root of the polynomial $(r_{\rm X}/{R^{3}_{\rm H}})r^{3}{+}r{-}r_{\rm X}{=}0$ with $r_{\rm X}{=}Gm_{\rm p}/2c^{2}_{\rm s,X}$. We use the \texttt{numpy.roots} function to evaluate $r_{\rm s}$. The Mach number $\mathcal{M}_{\rm X}(r){=}u(r)/c_{\rm s,X}$ also follows from equation \ref{equation:parker_tidal}; separating variables and integrating equation \ref{equation:parker_tidal} outwards from the sonic radius $r_{\rm s}$ yields for $u(r)/c_{\rm s,X}$ \citep{cra04}:

\begin{align}\label{equation:mach}
\begin{split}
    \mathcal{M}_{\rm X}(r)&=\sqrt{-W_{\rm -1}(-f(r))},\\
    f(r)&=\bigg(\frac{r}{r_{\rm s}}\bigg)^{-4}\exp\bigg[{4 r_{\rm X}\bigg(\frac{1}{r_{\rm s}}-\frac{1}{r}\bigg)-1+2\frac{r_{\rm X}(r^{2}_{\rm s}-r^{2})}{R^{3}_{\rm H}}}\bigg],
\end{split}
\end{align}

\noindent where $W_{\rm -1}$ is the $k{=}{-}1$ branch of the Lambert W function (appropriate for the supersonic portion of the flow through which we are integrating). We find that including the tidal term in equation \ref{equation:parker_tidal} shortens the sonic point distance so it lies inside the Hill sphere, consistent with numerical work \citep[][]{murchimur09}.

Armed with the velocity structure of the X-ray wind, we then determine whether X-ray or EUV drive evaporation by numerically integrating equation \ref{equation:tau_euv}. If equation \ref{equation:tau_euv} yields $\tau_{\rm EUV}{<1}$, we assume the flow is EUV-driven and take $L_{\rm PE}{=}L_{\rm XUV}$. Otherwise, we assume the flow is X-ray driven. We also assume the flow is X-ray driven if $r_{\rm s}{<}R_{\rm p}$, since in this case it will be launched at approximately the sound speed.

The stellar X-ray and EUV luminosity evolution follows:

\begin{align}\label{equation:Lxuvt}
\begin{split}
    L_{\rm X}(t)&=10^{-3.5}L_{\odot} \ \ \ [t{\leq}t_{\rm sat}] \\    
    &=10^{-3.5}\bigg(\frac{t}{t_{\rm sat}}\bigg)^{-1.21}L_{\star} \ \ [t{>}t_{\rm sat}]\\
    L_{\rm EUV}(t)&=\beta L_{\rm x}(t)(F_{\rm X}(t))^{\gamma},
\end{split}
\end{align}

\noindent where the X-ray flux $F_{\rm X}{=}L_{\rm X}/4\pi R^{2}_{\star}$ while $(\beta,\gamma){=}(176,-0.35)$ and $(3040,-0.76)$ for soft and hard EUV photons, respectively (\cite{kinwhe20}; we correct for a typo in $\beta$ for hard EUV as noted by \cite{karleetho25}). We add both soft and hard EUV photons to produce the total EUV flux. The X-ray power law exponent of $1.21$ was estimated by \cite{clashecoh12} \cite[see also][]{jacdavwhe12}, while $t_{\rm sat}{=}10^{8}$ yr \citep[e.g.][]{jacdavwhe12}.\footnote{Variations in $t_{\rm sat}$ are unlikely to affect our results quantitatively, as we begin our integrations well after stars have left saturation ($t{=}10^{9}$ yr vs. $t_{\rm sat}{\sim}10^{8}$ yr).} We set the stellar bolometric luminosity $L_{\star}{=}L_{\odot}$.

\subsubsection{RLO vs. PE}\label{subsubsection:rlovpe}

We determine when RLO transitions to PE by calculating the optical depth to high energy photons near the Roche surface. If the RLO outflow is optically thick to high energy radiation at L1, the gas will be heated by stellar bolometric emission up to the sonic point and PE cannot occur \citep[e.g.][]{laihelheu10}. Since X-rays penetrate deeper than EUV, the outflow will first become optically thin to X-rays, then to EUV (assuming X-rays are emitted by the host star). The optical depth to X-rays is,

\begin{equation}
    \tau_{\rm L1} \sim n_{\rm L1}\sigma_{\rm X}\ell_{\rm L1},
\end{equation}

\noindent where $n_{\rm L1}{=}\rho_{\rm L1}/\mu m_{\rm H}$, $\ell_{\rm L1}$ is the characteristic length scale of the flow near L1, and $\sigma_{\rm X}{=}1.2{\times}10^{-22}(Z/Z_{\odot}) \ \mathrm{cm}^{2}$ is the X-ray absorption cross section \cite[e.g.][]{nakhosyos18}. Since the outflow is subsonic below L1, the density distribution will be approximately hydrostatic \citep{lamcas99}:

\begin{equation}\label{equation:rhohse}
    \rho(r)\propto\exp\bigg({r_{\rm b}}\bigg[\frac{2}{r}-\frac{2}{R_{\rm p}}+\frac{r^{2}-R^{2}_{\rm p}}{R^{3}_{\rm H}}\bigg]\bigg),
\end{equation}
\noindent with $r_{\rm b}{=}G m_{\rm p}/2c^{2}_{\rm s}$ the Bondi radius. The scale height $\ell(r){=}\partial r/\partial \log P$ is then,

\begin{equation}\label{equation:lhse}
\ell(r)=\frac{c^{2}_{\rm s}}{g(r)}\frac{1}{1-(r/R_{\rm H})^{3}},
\end{equation}

\noindent with $g(r){=}GM_{\rm p}/r^{2}$. Because this expression for $\ell_{\rm L1}$ formally diverges at $r{=}R_{\rm H}$ (tidal gravity flattens the density distribution), we evaluate $\ell_{\rm L1}$ at 0.97${\times}R_{\rm H}$. In other words, to have a physically meaningful estimate of the optical depth to XUV photons at L1 during RLO, we cannot evaluate equation \ref{equation:lhse} at exactly $R_{\rm H}$; otherwise, equation \ref{equation:lhse} would always predict that RLO dominates over PE ($\tau_{\rm L1}{\rightarrow}\infty$) even when RLO shuts off. We have checked that the choice of where to evaluate $\ell_{\rm L1}$ does not significantly change our estimate for when RLO transitions to PE.

We adopt RLO when $\tau_{\rm L1}{\geq}1$. As RLO subsides due to orbital expansion during mass transfer (see section \ref{subsection:dynamics}), $\tau_{\rm L1}$ will drop and the flow may transition to PE-driven. Since PE does not involve mass transfer, the orbit will tidally decay while the planet evaporates. If tides are sufficiently strong, the decay could re-trigger an outburst of RLO. Thus, a duty cycle between RLO and PE may emerge with a timescale set by tidal decay. We smooth over this possible duty cycle by interpolating between RLO and photoevaporative mass loss using a sigmoidal function of $\tau_{\rm L1}$. We choose a sigmoid such that $\tau_{\rm L1}{=}1$ yields pure RLO, $\tau_{\rm L1}{=}0.1$ yields pure PE, and $\tau_{\rm L1}{=}0.5$ yields a mass loss rate half way between the two. We have verified that our results are robust against variations in the width/location of our smoothing function.

\begin{deluxetable*}{CCCCCCc}\label{param_table}
\tablecaption{Adopted calculation parameters.
\label{tab:parameters}}
\tablecolumns{6}
\tablewidth{0pt}
\tablehead{
\colhead{Parameter} &
\colhead{Definition} &
\colhead{Values} & 
\colhead{Units} & 
\colhead{References} &
}
\startdata
$m_{\rm p}$ & \rm initial \ planet \ mass & 3{\times}10^{2} & M_{\oplus} & \nodata \\
$m_{\rm c}$ & \rm \ planet \ core \ mass & $\{10,20\}$ & M_{\oplus} & \nodata \\
$M_{\star}$ & \rm \ stellar \ mass & 1.0 & M_{\odot} & \nodata \\
$Q_{\star}$ & \rm \ stellar \ tidal \ quality \ factor & 10^{$\{4,5,6\}$} & \nodata & \nodata \\
$X,Z$ & \rm atmospheric \ mass \ fractions & $\{0.74,0.02\}$, \ $\{0.36,0.5\}$ & \nodata & \nodata \\
$\Gamma$ & \rm angular \ momentum \ loss \ fraction & $\{\sqrt{R_{\star}/a},0.85\}$ & \rm \nodata & (1), \ (2) \\
$t_{\rm sat}$ & \rm XUV \ saturation \ time & 10^{8} & \rm yr & (3) \\
$L_{\rm X}(t{\leq}t_{\rm sat})$ & \rm saturated \ Xray \ luminosity & 10^{-3.5} & L_{\odot} & (3), (5) \\
$L_{\rm X}{\propto}t^{-\xi}$ & \rm Xray \ luminosity \ power \ law \ index & 1.21 & \nodata & (4) \\
($\beta$,$\gamma$) & \rm EUV/Xray \ ratio \ coefficient, \ power \ law \ index & (176,$-$0.35) (\rm soft), \ (3040,$-$0.76) (\rm hard) & \rm (erg \ s^{-1} \ cm^{-2}, \nodata) & (5) \\
\enddata
\tablerefs{(1) \cite{metgiaspi12}, (2) \cite{valrapras15}, (3) \cite{jacdavwhe12}, (4) \cite{clashecoh12}, (5) \cite{kinwhe20}}
\end{deluxetable*}

Our approach to treating RLO and PE is supported by the results of \cite{koslavhua22}, who argue that PE transitions to RLO once the EUV attenuation pressure \citep[i.e. where the atmosphere becomes optically thick to EUV photons, typically around ${\sim}$nBar;][]{murchimur09} lies beyond the Roche surface. Comparing to their Figure 6, we also find that PE activates at ${\sim}0.03$ au for a Uranus-like planet around a Sun-like star. Our approach differs from previous work \cite[][]{valrapras15} that imposed photoevaporation even when $R_{\rm p}{=}R_{\rm R}$ \cite[and therefore $\phi_{\rm L1}{=}\phi_{\rm ph}$; see also the discussion in][]{jacjenpea16}. We also note that our approach to determining when PE gives way to RLO also differs from that of \cite{jacarrpen17}, who estimate that PE only occurs once the Bondi radius lies inside the Hill sphere; our equation \ref{equation:mach} indicates that, if XUV photons can indeed penetrate the atmosphere, the sonic point will always lie inside L1 due to tidal acceleration. Our approach therefore produces more prevalent PE than that estimated by \cite{jacarrpen17}.

\subsection{Dynamical Evolution}\label{subsection:dynamics}

We follow the planetary orbital evolution under tidal decay due to tides raised on the star by the planet as well as angular momentum exchange during RLO. We ignore tides raised on the planet by the star as we assume planetary eccentricities and obliquities are zero. We adopt the equilibrium tide model of \cite{hut81}, integrating the following equations of motion due to dissipation in the star \citep{matpearas10}:

\begin{align}\label{equation:tides}
\begin{split}
\frac{\dot{a}_{\rm tid}}{a}&=-6 k_{2,\star}\delta t_{\star}\Omega\frac{m_{\rm p}}{M_{\star}}\bigg(\frac{R_{\star}}{a}\bigg)^{5}\bigg[\Omega-\omega_{\star}\bigg],\\
\frac{\dot{\omega}_{\star, \ \rm tid}}{\omega_{\star}}&=\frac{3}{\alpha_{\star}} k _{2,\star}\delta t_{\star}\Omega \frac{\Omega}{\omega_{\star}}\bigg(\frac{m_{\rm p}}{M_{\star}}\bigg)^{2}\bigg(\frac{R_{\star}}{a}\bigg)^{3}\bigg[\Omega-\omega_{\star}\bigg],
\end{split}
\end{align}

\noindent where $k_{2,\star}{=}0.028$ is the star's Love number (twice the apsidal motion constant of 0.014 for a Sun-like star; we note that $k_{2,\star}$ agrees within an order of magnitude across main sequence FGK stars and from solar to super-solar metallicities of [Fe/H]${=}0.5$ \citep{cla23}), $\delta t_{\star}{=}1/(2\Omega Q_{\star})$ is the tidal lag time, with $Q_{\star}$ the tidal quality factor \citep[\textit{not} the modified quality factor $Q^{'}_{\star}{=}3Q_{\star}/2k_{\rm 2,\star}$, e.g.][]{lai12}, $\omega_{\star}$ the stellar rotation frequency, and $\alpha_{\star}{=}0.06$ the square of the stellar radius of gyration \citep[the square of ``Beta" under Table II.1 from][]{cla89}. While current tidal theory predicts non-trivial relations between $\delta t_{\star}$, $\Omega$, and $Q_{\star}$ \citep[e.g.][]{ogi14}, we adopt this simplified formulation to showcase the basic physical mechanism operating during RLO. Future work should address how more sophisticated tidal physics alters the evolution of RLO. We initialize the star at a rotation period of 20 days, appropriate for ${\sim}$Gyr-old main sequence FGK stars \citep[e.g.][]{barweifri16}. In this work, we explore a broad range of stellar tidal quality factors $\log{Q_{\star}}{\in}\{4,5,6\}$ (see Figure \ref{figure:rlo_3_Qcomp} in Section \ref{sec:appendix_Q} for results with $\log{Q_{\star}}{=}6$).

Angular momentum transfer during RLO changes the planetary semimajor axis according to (for $q{\ll}1$),

\begin{equation}\label{equation:adot_rlo}
    \frac{\dot{a}_{\rm RLO}}{a}=-2\frac{\dot{m}_{\rm RLO}}{m_{\rm p}}\bigg[1-\Gamma\bigg],
\end{equation}

\noindent where the first term follows if the angular momentum carried away by the outflow is returned to the orbit. This return of angular momentum is thought to occur via tidal torque from an accretion disk formed under viscous settling of outflowing gas 
\citep[e.g.][]{ver82,priwad85}. Disk torques act oppositely to stellar tides if $\omega_{\star}{<}\Omega$, which is always the case in our calculations. 

The second term of equation \ref{equation:adot_rlo} arises if a fraction $\Gamma$ of the angular momentum carried away is not returned to the orbit. A useful parameterization of $\Gamma$ is to relate it to the radius $r_{\rm d}{=}\Gamma^{2}a$ at which outflowing gas settles into a disk/ring of angular momentum $\sqrt{GM_{\star}r_{\rm d}}$ before being lost from the system \citep[e.g.][]{sobphivan97}, or alternatively to write $\Gamma$ as the fraction of the angular momentum that is accreted directly onto the star \citep[``direct impact accretion", e.g.][]{marnelste04}.\footnote{$\Gamma$ also depends on the fraction $\delta$ of $\dot{m}_{\rm RLO}$ that settles into the disk/ring. For simplicity we set $\delta{=}1$ in this discussion without loss of generality.} Stellar accretion from the disk yields a lower limit of $\Gamma{=}\sqrt{R_{\star}/a}$, which is ${\sim}0.5$ at 0.02 au \citep[assuming a steady-state so that accretion occurs at the same rate as the planet transfers mass;][]{metgiaspi12}. An upper limit of $\Gamma{=}1$ follows if all the material leaving the planet is immediately lost. \cite{valrapras15} estimate intermediate values of $\Gamma{\sim}0.84{-}0.95$ for $m_{\rm p}{\sim}10{-}300\ M_{\oplus}$, consistent with the disk/ring radius estimated by \cite{laihelheu10} for the ${\sim}450 \ M_{\oplus}$ hot Jupiter WASP-12 b \citep[see e.g.][for observational evidence of a circumstellar torus/disk around WASP 12]{hasfosayr12}. We therefore explore $\Gamma{\in}\{\sqrt{R_{\star}/a},0.85\}$ as fiducial values throughout our calculation. Future numerical investigation is needed to fully characterize $\Gamma$; angular momentum loss could arise through magnetized winds/jets launched from the disk \citep[e.g.][]{bai16,nisantalc18}, outflows launched from the stellar surface as it spins up \citep[][]{matpud05,matabasta17,fraandbav23} and/or from the impact site between supersonic escaping gas and the disk surface \citep[see e.g.][for investigation of more exotic binary systems]{vandegdel08,dic24}. Angular momentum could also be lost by escape through L2, if the material flowing through L2 is lost from the system \citep[][]{metgiaspi12}. Such material could also exert an inward torque on the planet, either before escaping the system or by forming a disk external to the planet. We find however that flow through L2 can account for at most ${\sim}10\%$ of the total mass loss/angular momentum removed from the orbit during mass loss \citep[see e.g.][]{gulinbod03,laihelheu10}, so the effect of material leaving L2 is strongly subdominant to that at L1.

Our fiducial calculations do not include stellar spin up from accretion. We have verified that including spin up from accretion does not significantly alter the evolution since tidal spin up dominates. In Section \ref{subsec:strot} we do however explore how varying degrees of stellar spin up alter the predicted spread in stellar rotation rates from RLO.

\subsection{Summary of Calculation}

We simultaneously track the planetary thermal evolution using our pre-computed planetary structure grid and equation \ref{equation:cooling}, mass loss under RLO or PE using equations \ref{equation:mdot_rlo} and \ref{equation:mdot_pe} respectively (deciding between them as discussed in Section \ref{subsubsection:rlovpe}), as well as the semimajor axis evolution under tides and orbital expansion during RLO via equations \ref{equation:tides} and \ref{equation:adot_rlo}, respectively ($\dot{a}{=}\dot{a}_{\rm tid}{+}\dot{a}_{\rm RLO}$). The star spins up under tides via equations \ref{equation:tides}, while the XUV luminosity evolves via equations \ref{equation:Lxuvt}. The integration runs from $t{=}$1 to 10 Gyr, appropriate for treating a hot Jupiter orbiting a main sequence star. Planets are initialized at entropies $S{=}8$ $k_{\rm B}$/baryon (for $Z{=}0.02$) and $S{=}7$ $k_{\rm B}$/baryon ($Z{=}0.5$)  (approximately the largest $S$ at which we can find structure solutions across the entire range of masses and equilibrium temperatures we consider). 
Our parameter choices are summed up in Table \ref{param_table}.

Occasionally our integrations reach the boundary of our planet structure grid. This is usually triggered through total loss of atmosphere. We also find occasionally that planets with very little atmosphere fail to harbor convective interiors, rendering equation \ref{equation:cooling} inappropriate for treating their thermal evolution. In both cases, we simply shut off the atmospheric evolution and evolve the orbital dynamics.

\begin{deluxetable*}{CCCCCCc}\label{tab:fiducial_table}
\tablecaption{Structural parameters of our fiducial giant planet (used in Figure \ref{figure:evolution_metzger}). 
\label{tab:parameters}}
\tablecolumns{5}
\tablewidth{0pt}
\tablehead{
\colhead{Parameter} &
\colhead{Definition} &
\colhead{Value} & 
\colhead{Units} & 
}
\startdata
$m_{\rm p}$ & \rm initial \ planet \ mass & 3{\times}10^{2} & M_{\oplus} \\
$m_{\rm c}$ & \rm \ planet \ core \ mass & 10 & M_{\oplus} \\
$Q_{\star}$ & \rm \ stellar \ tidal \ quality \ factor & 10^5 & \nodata \\
$X,Z$ & \rm atmospheric \ mass \ fractions & $\{0.74,0.02\}$ & \nodata \\
$\Gamma$ & \rm angular \ momentum \ loss \ fraction & $\sqrt{R_{\star}/a}$ & \nodata \\
\\
\enddata
\end{deluxetable*}

\section{Results}\label{sec:results}

Our goal in Section \ref{subsec:standard} is to understand the outcome of Roche lobe overflow in the case of low angular momentum loss, which we refer to as ``standard" RLO. Under standard RLO, angular momentum is lost solely through stellar accretion from the disk, so that $\Gamma{=}\sqrt{R_{\star}/a}$. We begin by considering the fate of a fiducial gas giant (parameters for our fiducial gas giant are given in Table \ref{tab:fiducial_table}), and highlight how standard RLO does not produce sufficient mass loss to reproduce the desert dwellers. In Section \ref{subsec:lossy} we then contrast standard RLO with RLO under high degrees of angular momentum loss (``lossy" RLO; $\Gamma{=}0.85$). We first articulate the effect of varying core mass, then in Section \ref{subsec:parameters} outline how atmospheric metallicity, the strength of tides, and entropy also dictate the outcome of lossy RLO. We will show that desert dwellers robustly emerge in all cases of lossy RLO. 

\subsection{``Standard" RLO}\label{subsec:standard}

\begin{figure}
\epsscale{1.26}
\plotone{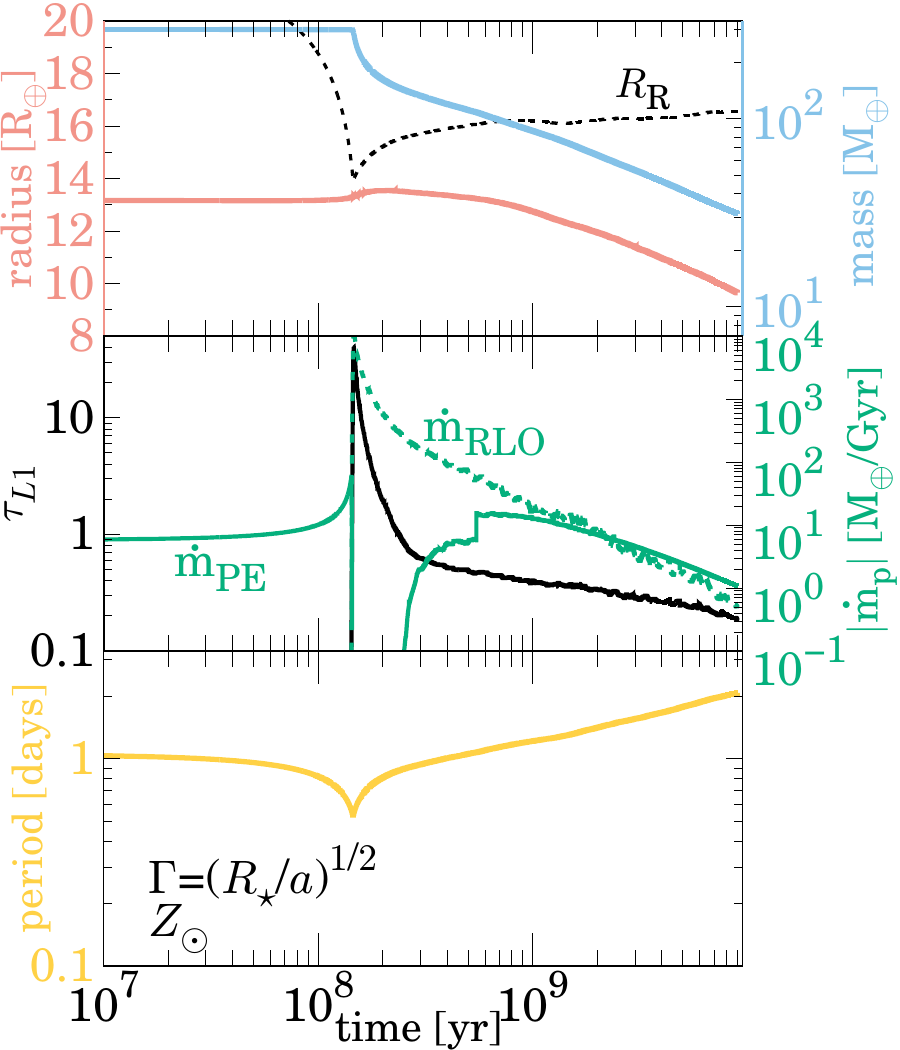}
\caption{Evolution of a fiducial giant planet (see Table \ref{tab:fiducial_table} for planetary structural parameters) under standard angular momentum loss ($\Gamma{=}\sqrt{R_{\star}/a}$). Top panel: radius (red; left axis), Roche radius (black), and mass (blue; right axis) through time. Middle panel: optical depth to XUV photons at L1 (black; left axis) and mass loss rates (photoevaporation in solid green, RLO in dashed; right axis). Bottom panel: orbital period. Tidal decay triggers RLO before the planet bounces out to larger periods as angular momentum is returned to the orbit. Roche lobe overflow removes ${\sim}90\%$ of the planet's mass, while photoevaporation plays a minor role (the uptick in $\dot{m}_{\rm PE}$ at ${\sim}6{\times}10^{8}$ yr occurs as PE becomes dominated by UV rather than X-ray radiation). Despite its violent mass loss history, the planet remains puffy and harbors a substantial gas envelope. \label{figure:evolution_metzger}}
\end{figure}

In Figure \ref{figure:evolution_metzger} we display the evolution of a fiducial giant planet under standard RLO (structural parameters for our fiducial giant planet are provided in Table \ref{tab:fiducial_table}). Orbital decay under stellar tidal dissipation initially causes $R_{\rm R}$ to decrease until the planet radius lies within ${\lesssim}1 \ R_{\oplus}$ of the Roche radius (from $t{=}10^{7}$ to $t{\sim}1.5{\times}10^{8}$ yr). The shallow potential well at this moment of maximum Roche lobe contact (at $t{\sim}1.5{\times}10^{8}$ yr), coupled with the hot equilibrium temperature at such short periods (${\sim}2500$ K), catalyzes intense hydrodynamic escape through L1. Photoevaporation gives way to RLO as the optical depth to XUV at L1 surges. Orbital expansion in response to rapid mass transfer then damps the mass transfer rate by expanding $R_{\rm R}$ (beginning at $t{\gtrsim}1.5{\times}10^{8}$ yr). Mass transfer and orbital expansion conspire to further pare down the mass transfer rate for the remainder of the evolution; lower planet masses compensate for lower mass loss rates (equation \ref{equation:adot_rlo}; $\dot{a}_{\rm RLO}{\propto}|\dot{m}_{\rm p}|/m_{\rm p}$) to continually push the planet further out where mass loss rates continue to decline ($\dot{a}_{\rm RLO}{\propto}a$). Planetary cooling contraction also contributes to decreasing the mass loss rate as the planet ages and moves outward. A second period of photoevaporative mass loss occurs once the RLO outflow is sufficiently weak ($t{\gtrsim}2{\times}10^{9}$ yr), however the planet remains with a substantial amount of gas and the radius of a giant planet.

\begin{figure*}
\centering
\includegraphics[width=0.9\textwidth]{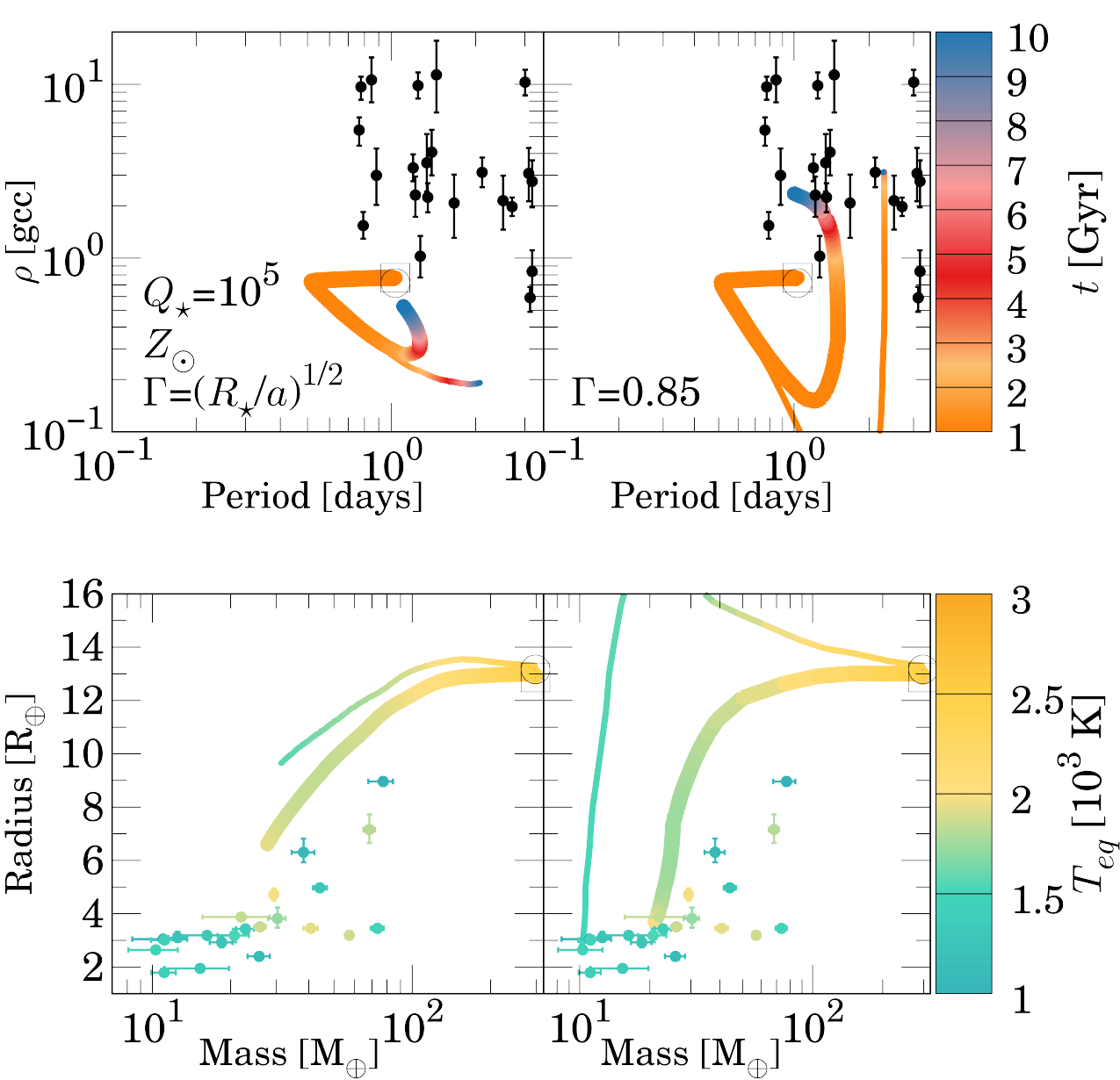}
\caption{\textit{Top:} Evolution of planet density versus orbital period through time (color bar; circular and square points demark initial conditions for the 10 and 20 $M_{\oplus}$ cores) compared to that of observed desert dwellers (data points), for low angular momentum loss (${\Gamma}{=}\sqrt{R_{\star}/a}$; left panels) and high ($\Gamma{=}0.85$; right panels). Evolution tracks illustrated in thin lines use a 10 $M_{\oplus}$ core, while those in thick lines use a 20 $M_{\oplus}$ core. \textit{Bottom:} Planetary mass, radius and equilibrium temperature (color bar) as compared to observed desert dwellers. If angular momentum loss from the orbit is weak, planets emerge from RLO underdense compared to the observations. If RLO is ``lossy" (i.e. the majority of angular momentum lost from the planet by the outflow is not returned to the orbit), gas giants successfully shed ${\sim}99\%$ of their atmosphere. Their remnants boast densities in line with those observed.
\label{figure:mrcomp}}
\end{figure*} 

The lesson of Figure \ref{figure:evolution_metzger} is that, in the absence of significant angular momentum loss (i.e. with $\Gamma{=}\sqrt{R_{\star}/a}$), RLO cannot achieve the ${\sim}99\%$ atmospheric removal needed to produce a planet with density similar to those in the desert. This shortcoming of RLO in producing planets that are underdense compared to those observed in the desert is underscored in the left panels of Figure \ref{figure:mrcomp}. Figure \ref{figure:mrcomp} also illustrates that using a more massive core, and therefore denser model planets, still cannot bring model planets' densities into alignment with those observed. We note that such low planetary densities are similar to those previously found using independent methods to calculate the evolution \cite[see Figure 5 of][]{valrapras15}.

\begin{figure*}
\centering
\includegraphics[width=1\textwidth]{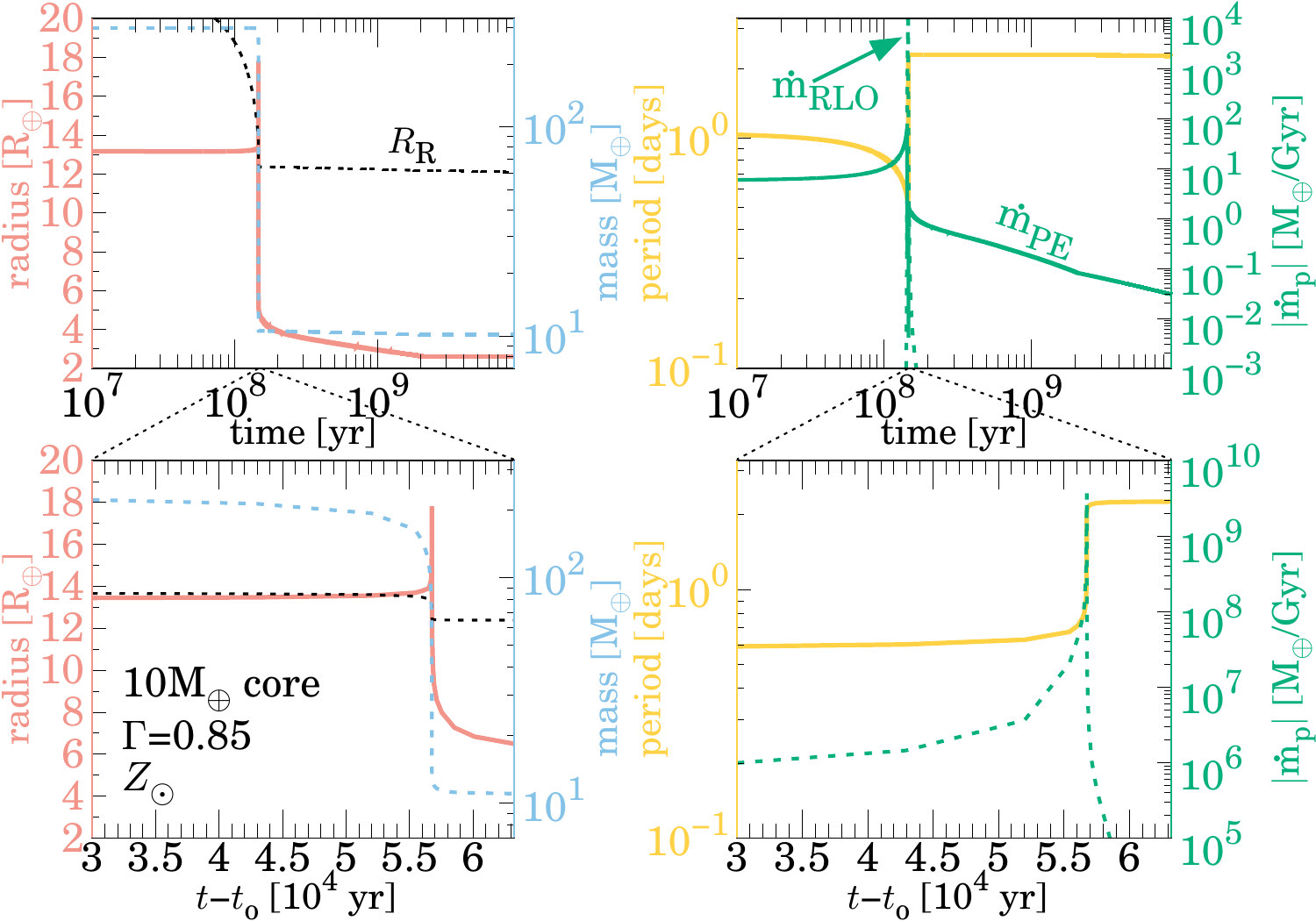}
\caption{Evolution of a fiducial giant planet under ``lossy" RLO (i.e. 85$\%$ of the angular momentum lost from the planet during mass transfer is not returned to the orbit). Top panels depict the full evolution, while bottom panels zoom in near the time of runaway RLO. Red axes correspond to radius, blue to mass, yellow to orbital period, and green to mass loss rate. Lossy RLO yields runaway mass loss whereby a giant planet loses ${\sim}99\%$ of its atmosphere. The remaining atmosphere passively photoevaporates to destruction, revealing a stripped core in the sub-Jovian desert.
\label{figure:evolution_zoom}}
\end{figure*} 

\begin{figure}
\epsscale{1.26}
\plotone{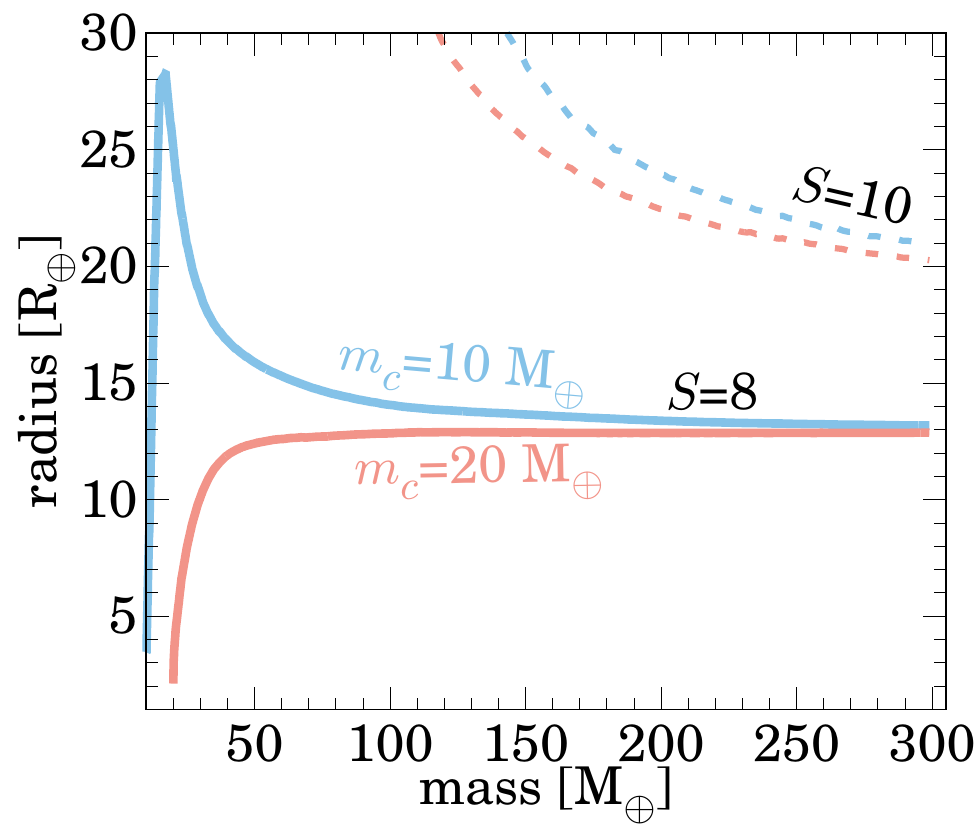}
\caption{Adiabatic mass-radius relations for a sample of planets from our grid. Solid and dashed curves denote planets with internal entropies of $S{=}8$ and 10 $k_{\rm B}/m_{\rm H}$, respectively. Blue curves use a 10 $M_{\oplus}$ core, while red use 20 $M_{\oplus}$. Planets' adiabatic mass-radius relations dictate the evolution during runaway RLO. \label{figure:mr}}
\end{figure}

\begin{figure*}
\centering
\includegraphics[width=1\textwidth]{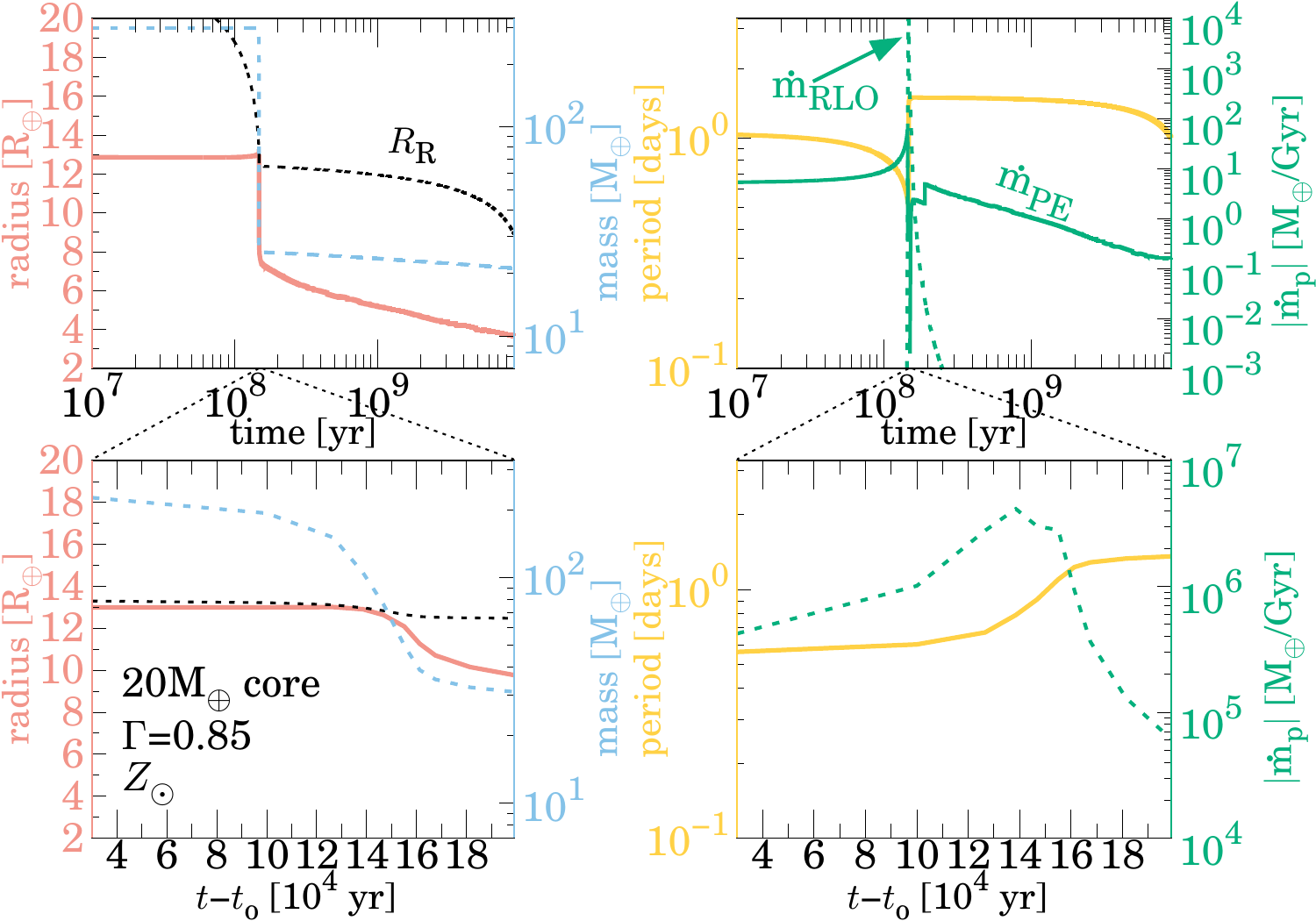}
\caption{Evolution of a giant planet with a more massive core than our fiducial case from Figure \ref{figure:evolution_zoom}. Both the planet from Figure \ref{figure:evolution_zoom} and this figure begin with the same initial entropy. Despite the fact that planets with more massive cores do not undergo adiabatic radius inflation during mass loss (at our chosen entropy), lossy RLO still evacuates ${\sim}96\%$ of the atmosphere. The remaining atmosphere evaporates to destruction over Gyr.\label{figure:evolution_mc=20}}
\end{figure*} 

\

\subsection{``Lossy" RLO}\label{subsec:lossy}

The right panels of Figure \ref{figure:mrcomp} however show that the picture changes in the presence of significant angular momentum loss during RLO ($\Gamma{=}0.85$): in this case, gas giants successfully shed the vast majority of their gas to emerge as thinly veiled cores in the desert, with similar sizes to those observed. The reason for the change of mass loss behavior is outlined in Figure \ref{figure:evolution_zoom}. Unlike the fiducial case from Figure \ref{figure:evolution_metzger}, ``lossy" RLO yields runaway mass loss. The runaway arises due to the fact that in the presence of significant angular momentum loss, the orbit cannot immediately expand and reduce $|\dot{m}_{\rm RLO}|$ in response to mass transfer. As a result, tidal decay continues to shrink the orbit until the planet comes into contact with the Roche lobe ($R_{\rm p}{\gtrsim}R_{\rm R}$). Overflowing the Roche lobe yields adiabatic mass loss; the planet cannot cool in the time it takes to lose mass ($S{/}|\dot{S}|{\gg}m_{\rm p}{/}|\dot{m}_{\rm p}|$), so the evolution after the moment of Roche lobe contact depends on the planet's adiabatic mass-radius relationship. We display example adiabatic mass-radius relationships in Figure \ref{figure:mr} (the solid blue curve corresponds to the planet shown in Figure \ref{figure:evolution_zoom}). As displayed in both Figures \ref{figure:evolution_zoom} and \ref{figure:mr}, the planet's radius \textit{grows} as it loses mass at fixed entropy, further increasing the mass loss rate (see \cite{barselcha04} and \cite{kurnak14} for related discussion in the context of photoevaporation, though we note that photoevaporation is understood to be ineffective for giant planets; \cite{murchimur09,owejac12}).\footnote{We note that our structure calculations do not account for $PdV$ work exerted as the atmosphere undergoes intense mass loss, unlike the calculations presented in \cite{barselcha04}. Including $PdV$ work would \textit{exacerbate} the mass loss induced by lossy RLO; as reported by \cite{barselcha04}, rapid shedding of radiative upper layers can produce $PdV$ radius expansion, making the planet even more susceptible to mass loss.} This increase in $|\dot{m}_{\rm p}|$ is exacerbated by the fact that the Roche radius \textit{declines} during RLO as the planet's lower mass outweighs the increasing orbital period. The result of this feedback is almost wholesale destruction of the gas giant's atmosphere.

However, Figure \ref{figure:mr} also shows that the adiabatic mass-radius relation differs for planets with more massive cores: the radius of a planet with a 20 $M_{\oplus}$ core is almost constant down to masses ${\lesssim}50 \ M_{\oplus}$, below which radius declines (at low entropy $S{=}8 \ k_{\rm B}/m_{\rm H}$). As shown in Figure \ref{figure:evolution_mc=20}, such a mass-radius relation produces lower maximum mass loss rates (due to the absence of adiabatic radius inflation) but nevertheless the planet sheds the vast majority of its envelope via RLO. Figure \ref{figure:evolution_mc=20} therefore highlights the fact that \textit{adiabatic radius inflation is not necessary to induce runaway RLO}; what \textit{is} necessary is that RLO be lossy. Regardless of the core mass, runaway RLO damps once the planets have lost so much mass that they cannot physically fill the Roche radius. Mass transfer (at fixed entropy) therefore terminates at larger masses for planets with more massive cores, due to the downturn in their mass-radius relation.

\subsection{Dependence on Loss Fraction $\Gamma$}

In this section our goal is to flesh out the range of angular momentum loss fractions $\Gamma$ that successfully produce desert dwellers. Rerunning our calculations with $\Gamma{\in}\{0.65,0.75,0.95\}$ reveals that $\Gamma{\gtrsim}0.95$ yields immediate tidal engulfment, $\Gamma{\lesssim}0.65$ yields underdense planets similar to Figure \ref{figure:evolution_metzger}, while $0.65{\lesssim}\Gamma{\lesssim}0.95$ generally reproduces our desert dwellers. We thus conclude that a broad range of $\Gamma$ can induce sufficiently violent RLO to form desert dwellers.

\subsection{How Metallicity, the Strength of Tides, and Entropy Affect Lossy RLO}\label{subsec:parameters}

\begin{figure*}
\centering
\includegraphics[width=1\textwidth]{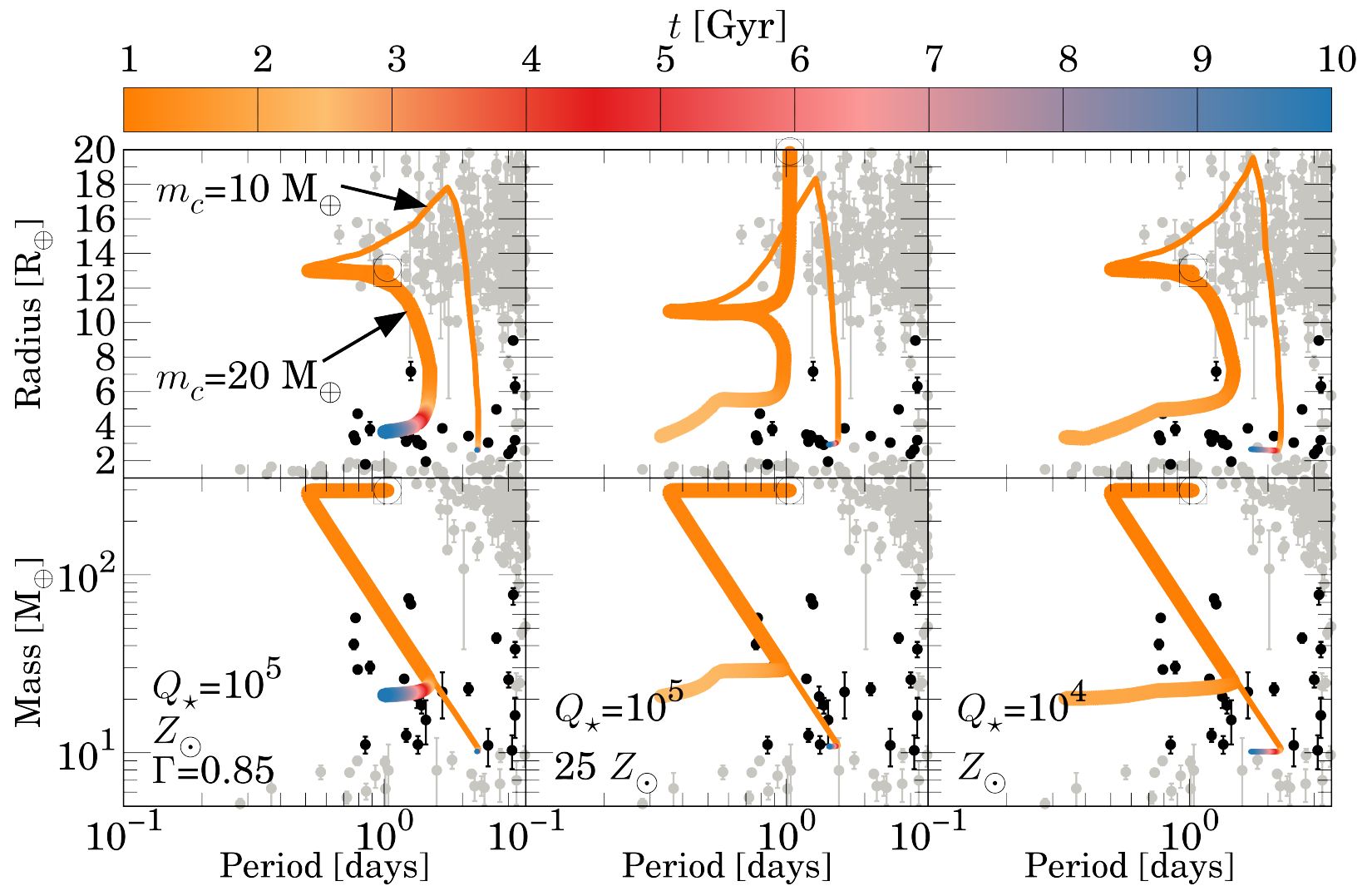}
\caption{Exploring the effects of metallicity and tidal efficiency on the evolution tracks of model planets under lossy RLO. Solar metallicity planets with $Q_{\star}{=}10^{5}$ are shown in the left panels, metal-rich planets in the middle, and those under enhanced stellar tides ($Q_{\star}{=}10^{4}$, solar metallicity) in the right panels. Thin lines correspond to models with a 10 $M_{\oplus}$ core (initial condition demarcated by the circular data point), thick lines to those with a 20 $M_{\oplus}$ core (initial condition demarcated by the square data point). Planets that emerge via lossy RLO spend ${\sim}$Gyr in the desert with similar radii and masses to those observed (black data). Metal-rich desert dwellers are more prone to tidal decay and generally emerge at shorter periods than their lower metallicity counterparts. The strength of stellar tides changes the long-term inspiral behavior but not the mass loss evolution.
\label{figure:rlo_3}}
\end{figure*}

Figure \ref{figure:rlo_3} explores the effects of metallicity and the strength of stellar tides on the evolution. First, we observe the fact that planets that form via lossy RLO generally spend Gyr in the desert with similar densities to those observed. It is also clear from Figures \ref{figure:mrcomp} and \ref{figure:rlo_3} that planets with larger cores end their mass transfer evolution at shorter periods than those with less massive cores. As discussed above, this trend is tied to the planets' adiabatic mass-radius relationships. Planets with lower core masses experience higher maximum mass loss rates due to adiabatic radius inflation ($\dot{a}_{\rm RLO}{\propto}{-}\dot{m}_{\rm RLO}$), in addition to the fact that they continue mass transfer, and therefore replenish their orbital angular momentum, down to lower masses because their radii remain inflated ($\dot{a}_{\rm RLO}{\propto}1/m_{\rm p}$). 

Figure \ref{figure:rlo_3} also reveals that metal-rich planets end their evolution at shorter periods than their low-$Z$ counterparts. This trend is partly due to the fact that since high-$Z$ planets are denser (they hunker down), RLO is initiated at smaller orbital periods where tides are stronger. Even after initiating RLO however, metal-rich planets can exhibit substantially lower mass loss rates (at a given period) owing to the high opacity of their atmospheric constituents decreasing the photospheric density, as well as sound speed via elevated mean molecular weight. Lower photospheric density and sound speed directly reduce the mass loss rate through equations \ref{equation:mdot_rlo} and \ref{equation:rho_roche}. Their reduced mass loss rates dictate that RLO shuts off earlier in the mass transfer process, further leaving planets more susceptible to tidal decay (photoevaporation is also quenched due to enhanced cooling in the metal-rich gas; equation \ref{equation:mdot_pe}). These effects are exemplified by the high-$Z$ planet with a 20 $M_{\oplus}$ core shown in Figure \ref{figure:rlo_3}; it emerges from the initial episode of RLO still harboring a substantial gas envelope before entering a second period of RLO in the midst of tidal inspiral. In the process, it boasts a density similar to the observed desert dwellers at periods $<$1 day, e.g. LTT 9779 b (see Section \ref{subsec:ltt9779} for detailed discussion).

The rightmost panel of Figure \ref{figure:rlo_3} shows that, as expected, stronger stellar tides produce earlier inspiral and RLO. By limiting the amount of time planets can cool before triggering RLO, stronger tides manifest as more pronounced adiabatic radius inflation for the planets with 10 $M_{\oplus}$ cores. However, varying $Q_{\star}$ only appreciably changes the long-term tidal decay of the system and not the mass loss evolution (see also Figure \ref{figure:rlo_3_Qcomp}). We also note that planetary evolution plays out in a similar way between that of metal-rich planets with weaker tides and low-$Z$ planets under strong tides due to the fact that the dynamics of both are dominated by tidal decay; high-$Z$ planets suffer lessened mass loss rates and so remain susceptible to tidal decay, compensating for their weaker $Q_{\star}$.

\begin{figure*}
\centering
\includegraphics[width=1\textwidth]{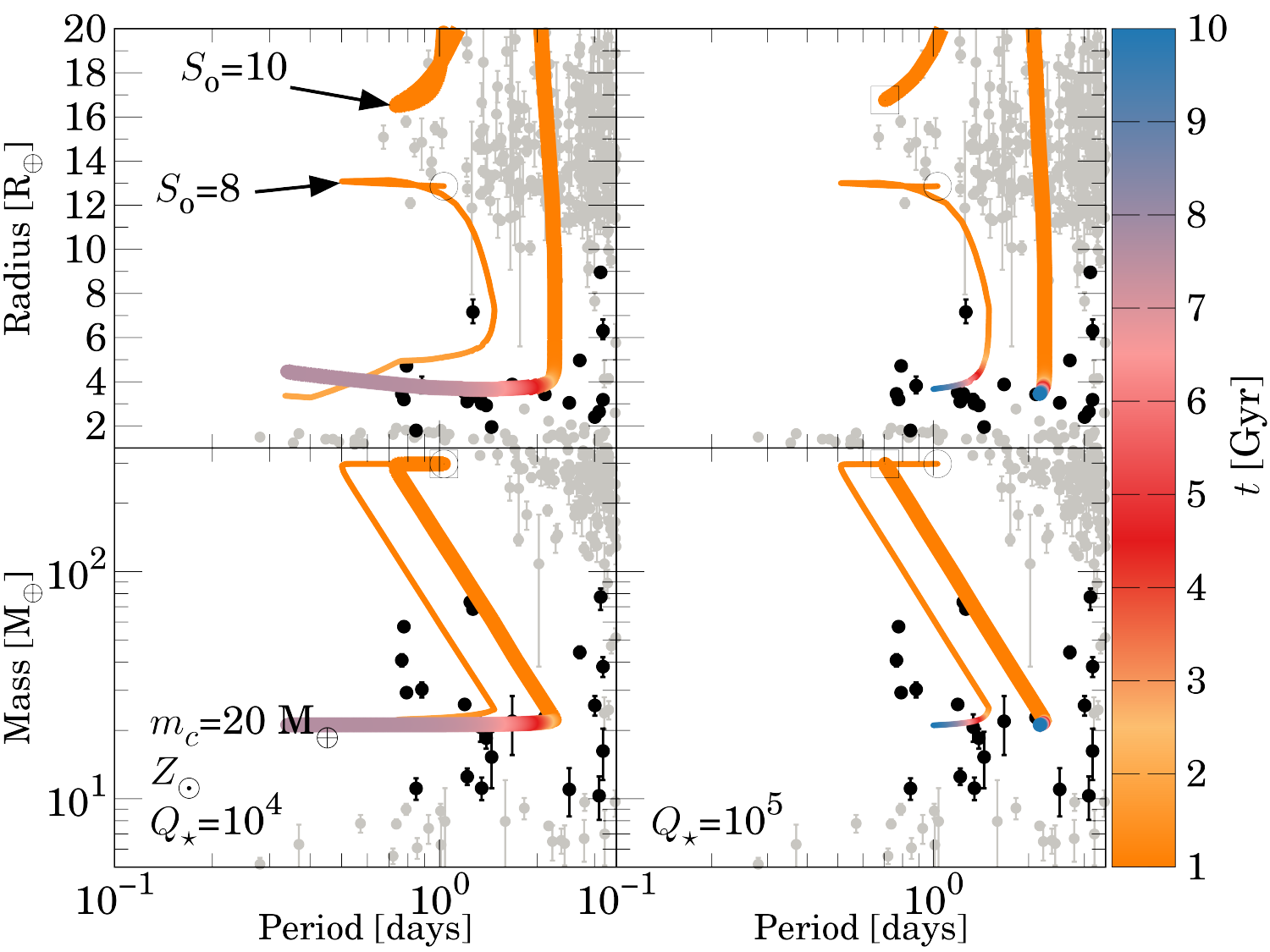}
\caption{Evolution tracks for planets with high initial entropy ($S_{o}{=}10$ $k_{\rm B}/m_{\rm H}$; thick curves) and low ($S_{o}{=}8$ $k_{\rm b}/m_{\rm H}$; thin curves) under strong stellar tides ($Q_{\star}{=}10^{4}$; left panels) and weak ($Q_{\star}{=}10^{5}$; right panels). All planets contain a 20 $M_{\oplus}$ core. Initial entropy and the strength of stellar tides are partially degenerate; observed desert dwellers at periods ${\sim}$1 day could form under weak tides and low initial entropy, or vice versa. The entire width of the sub-Jovian desert out to ${\lesssim}3$ days can be backfilled with the remnants of hot Jupiters that possessed their empirically inferred spread in entropy before destruction \cite[$S{\sim}8{-}11$ $k_{\rm B}/m_{\rm H}$; e.g.][]{thogaofor19}.  \label{figure:evolution_scomp}}
\end{figure*}

Figure \ref{figure:evolution_scomp} contrasts the evolution between our fiducial planets and those that begin with higher entropy. Since hot Jupiters harbor a range of internal entropies ${\sim}8{-}11$ $k_{\rm B}/m_{\rm H}$ \citep[e.g.][]{thogaofor19}, we chose $S_{o}{=}10$ $k_{\rm B}/m_{\rm H}$ as a representative value to compare our fiducial planets against. A higher initial entropy yields earlier initiation of RLO during tidal decay due to the planets' inflated radii, and promotes adiabatic radius inflation during mass loss (see Figure \ref{figure:mr}; the radii during RLO are so large that they exceed the axes of Figure \ref{figure:evolution_scomp}). As a result of the larger initial orbital period where RLO begins, as well as the higher mass loss rates afforded by adiabatic radius inflation, RLO takes place at larger orbital periods for planets at higher entropies. The upshot of Figure \ref{figure:evolution_scomp} is that the entire width of the desert out to ${\lesssim}3$ days can be backfilled by the remnants of hot Jupiters that began RLO with their observed spread in entropy. Unlike previous work \citep{valrapras15}, we find that top-down formation of desert dwellers therefore does not predict a correlation between planet mass and orbital period (i.e. marginalizing across initial entropy would likely erase any correlation between planet mass and final orbital period).

Figure \ref{figure:evolution_scomp} also underscores the fact that the stellar tidal quality factor and the initial entropy are partially degenerate; e.g. ${\sim}20 \ M_{\oplus}$ desert dwellers at periods ${\sim}1$ day (of which there are several) could arrive at their periods via strong tidal decay of destroyed high entropy gas giants (thick curve, left panel), or could emerge in-situ via RLO of a lower entropy progenitor under weak tides (thin curve, right panel). Our theory on its own therefore cannot determine the necessary strength of stellar tides to reproduce some of the observed desert dwellers. 

\section{Discussion}\label{sec:discussion}

\subsection{Forming LTT 9779 b $\&$ TOI 3261 b}\label{subsec:ltt9779}

In Section \ref{sec:results} we noted the similarity between the remnants of metal-rich gas giants and the closest-in desert dwellers at periods $<$1 day. The purpose of this section is to verify whether the top-down formation mechanism explored in this work can indeed exactly reproduce the sizes and periods of the hottest desert dwellers, and if so, extract theoretical predictions that can be used to test whether they formed in a top-down fashion. We focus specifically on LTT 9779 b as a test case \cite[$m_{\rm p}{=}29.32^{+0.78}_{-0.81} \ M_{\oplus}$, $R_{\rm p}{=}4.72^{+0.23}_{-0.23} \ R_{\oplus}$, orbital period 0.792 days;][]{jendiamat20}, since it boasts the highest equilibrium temperature of all desert dwellers \citep[1978${\pm}$19 K;][]{jendiamat20} and is arguably the most well-studied ultrahot Neptune. We expect the qualitative conclusions that we draw from studying the possible formation history of LTT 9779 to also apply to the other ultrahot Neptunes in the desert. Specifically, we will find that to simultaneously reproduce their short orbital periods and high densities requires massive cores, high metallicity atmospheres ($Z{\sim}0.5$), and ongoing present-day tidal decay (under the assumption of constant stellar tidal quality factors).

\begin{figure}
\epsscale{1.26}
\plotone{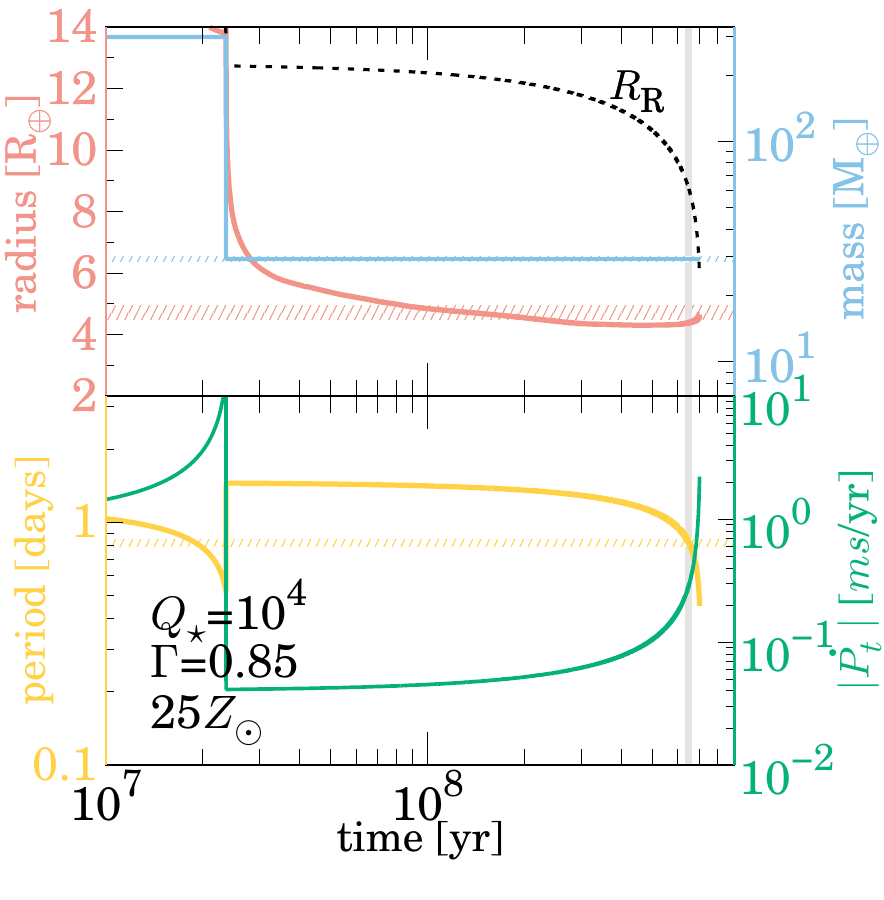}
\caption{Possible formation pathway of LTT 9779 b via tidally-driven Roche destruction of a giant planet. Present-day radius (red), mass (blue), and orbital period (yellow) for LTT 9779 b are shown as dashed bands encompassing their uncertainties following the measurements from \cite{jendiamat20} (the orbital period band has been made artificially wide to aid visualization). The vertical grey band demarcates the time when LTT 9779 b is recovered. Top-down formation favors a supersolar metallicity atmosphere for LTT 9779 b, in line with current measurements. If LTT 9779 b is the remnant of a giant planet, it should currently be tidally decaying at a rate ${\sim}0.5$ ms yr$^{-1}$. \label{figure:evolution_ltt}}
\end{figure}

As displayed in Figure \ref{figure:evolution_ltt}, we are able to reproduce LTT 9779 b's mass, radius, and period by considering a planet with a 23 $M_{\oplus}$ core and a metal-rich atmosphere that undergoes RLO. A slightly more massive core than that used in our fiducial models is necessary to reproduce LTT 9779 b's large density, since the planet undergoes approximately the same amount of mass loss as the high-$Z$ planet with a 20 $M_{\oplus}$ core shown in Figure \ref{figure:rlo_3} but stays more compact at the same entropy and mass. LTT 9779 b's high density similarly favors a metal-enhanced atmosphere. A metal-rich composition also helps LTT 9779 b keep what little atmosphere remains post-RLO by resisting photoevaporation. In this scenario, LTT 9779 b is currently evaporating at a rate $|\dot{m}_{\rm PE}|{\sim}2{\times}10^{10}$ g s$^{-1}$. Given our estimate for the planet's present day gas to core mass ratio GCR${=}26\%$, the current evaporation rate yields a mass loss timescale $\mathrm{GCR}{\times}m_{\rm p}/|\dot{m}_{\rm p}|{\gtrsim}10^{10}$ yr ($\dot{m}_{\rm RLO}{=}0$ in the present day). This estimate for $\dot{m}_{\rm p}$ is likely an upper limit as the host star exhibits an anomalously low X-ray luminosity \citep[][]{ferwhekin24}, but is consistent with current null detections of a planetary outflow (e.g. \cite{edwchatsi23,ramjensed25}; see also \cite{vismccdos24} who estimate an upper limit on $|\dot{m}_{\rm p}|$ similar to ours). Our fiducial metallicity of $25 \ Z_{\odot}$ is also consistent with that estimated from spectral analysis \citep[][]{radcoutay24,ramjensed25}, phase curve observations \citep[][]{crodracow20} and secondary eclipse measurements \citep[][]{hoyjenpar23}.

Recovering LTT 9779 b with our top-down formation theory necessitates a metal-rich atmosphere and a massive core. As discussed in Section \ref{sec:results}, the value of $Q_{\star}$ required to reproduce LTT 9779 b with our theory is degenerate with the initial entropy of the progenitor gas giant. We can however place a limit on the value of $Q_{\star}$ such that LTT 9779 b's progenitor gas giant tidally decayed to the point of RLO by the star's current age of $t_{\star}{=}2^{+1.3}_{-0.9}$ Gyr (\cite{edwchatsi23}; see also \cite{jendiamat20} who estimate an age of 1.9$^{+1.7}_{-1.2}$ Gyr). From equation \ref{equation:tides} we set the tidal decay timescale $t_{\rm tid}{\sim}2a/(13|\dot{a}|){=}t_{\star}$ \cite[including the 2/13 prefactor from][which assumes constant $Q_{\star}$]{barogi09} and solve for $Q_{\star}$ to yield (assuming $P{\ll}P_{\star}$),

\begin{align}\label{equation:ltt_Q}
\begin{split}
Q_{\star}&=\frac{78\pi}{2}k_{2,\star}\frac{t_{\star}}{P}\frac{m_{\rm p}}{M_{\star}}\bigg(\frac{R_{\star}}{a}\bigg)^{5}\\
         &{\sim}10^{4}\bigg(\frac{t_{\star}}{3 \ \rm Gyr}\bigg)\bigg(\frac{m_{\rm p}}{300 \ M_{\oplus}}\bigg)\bigg(\frac{P}{4 \ \rm days}\bigg)^{-13/3},
\end{split}
\end{align}

\noindent where we have used a solar-type star, appropriate for LTT 9779 \citep[which has mass $1.02^{+0.02}_{-0.03} \ M_{\odot}$ and radius $0.949{\pm}0.006 \ R_{\odot}$;][]{edwchatsi23} and $P$ corresponds to the orbital period of LTT 9779 b's progenitor before tidal decay and RLO. Equation \ref{equation:ltt_Q} therefore indicates that tidal quality factors $Q_{\star}{\gtrsim}10^{4}$ (modified quality factors $Q^{'}_{\star}{=}3Q_{\star}/2k_{2,\star}{\gtrsim}5{\times}10^{5}$) are likely too large to allow a hot Jupiter at a typical orbital period ${\sim}$4 days \citep[e.g.][]{gauseamal05,petmarwin18} to initiate RLO by the age of LTT 9779 b. If LTT 9779 b is the remnant of a tidally stripped gas giant, our theory therefore predicts that it should presently be undergoing tidal decay at a rate ${\sim}0.5$ ms yr$^{-1}$ (with a tidal inspiral time $P/|\dot{P}|{\sim}10^{8}$ yr).\footnote{We also find that in order to match LTT 9779 b’s radius and period in the present day, the tidal decay rate cannot be significantly smaller than this approximate lower bound. Significantly smaller tidal decay rates produce two problematic outcomes: either they strand high-entropy planets at large periods after RLO (shortly after $2{\times}10^{7}$ yr in Figure \ref{figure:evolution_ltt}), or planets hardly bounce back out to larger periods during RLO, owing to prolonged cooling during slow tidal decay toward RLO.} While such a decay rate is not detectable given current baselines \citep[][]{edwchatsi23}, future observations may be leveraged to pin down $\dot{P}$. We note that this estimate for $\dot{P}$ assumes a constant $Q_{\star}$, while in reality stellar tidal quality factors may vary with planetary orbital period \citep[e.g.][]{milmac25}. Future work extending our RLO calculations with more sophisticated tidal physics could be helpful to make more accurate tidal decay rate predictions.

The top-down formation of LTT 9779 b presented in this section involves tidal spin-up of the host star (see Section \ref{subsec:strot}). LTT 9779 however exhibits weak measured rotational broadening, which could reflect an exceptionally slow rotation velocity \citep[][]{jendiamat20}. A sluggish stellar rotation rate for LTT 9779 may therefore stand in tension with our picture. Since the measured stellar rotation velocity depends on both the true rotation velocity as well as the direction along which the rotation axis points relative to Earth (via $v\sin i$), stellar rotation may serve as a test of our theory only at a population level where the effect of viewing geometry in individual systems may be lessened. If LTT 9779 indeed exhibits slow rotation as a result of a spin axis oriented perpendicularly to the plane of the sky, we would therefore expect LTT 9779 b's orbital angular momentum axis to be strongly misaligned with the stellar spin axis to maintain transiting geometry (such a near-polar obliquity may for example be generated via inertial wave dissipation in the stellar interior, which can lock planets into polar configurations over the age of LTT 9779; \citealt{lai12}.) Follow-up observations of LTT 9779 b's stellar obliquity could therefore shed light on its origin channel and the anomalous stellar rotation rate. An alternative possibility is that LTT 9779's high metallicity facilitates exceptionally fast post-mass transfer spin down via wind-braking \citep[][]{amamat20}.

A similar picture can also account for TOI 3261 b, whose mass and radius almost overlap within their error bars with LTT 9779 b \cite[$30.3^{+2.2}_{-2.4} \ M_{\oplus}$, $3.82^{+0.42}_{-0.35}  \ R_{\oplus}$, with an orbital period and age of 0.88 days and 6.5${\pm}2.1$ Gyr respectively;][]{nabhuabur24}. Since the planet parameters for TOI 3261 b are so similar to those of LTT 9779 b, we opt not to include a detailed analysis of its possible evolutionary histories here but fully expect the results of such work to strongly resemble that of Figure \ref{figure:evolution_ltt} (i.e. top-down formation will require a metal-rich atmosphere and massive ${\sim}20 \ M_{\oplus}$ core). Unlike LTT 9779 however, TOI 3261 rotates more rapidly with a period ${\lesssim}16$ days \citep[][]{nabhuabur24}, which remains consistent with post-RLO spin down (see Figure \ref{figure:st_spin}). We therefore do not predict that TOI 3261 b possesses a non-zero stellar obliquity. We note that, contrary to the claim of \cite{nabhuabur24}, \cite{valrapras15} do not appear to recover TOI 3261 b or LTT 9779 b but produce planets that are underdense and at orbital periods that are too small (see their Figures 5 and 6, where 30 $M_{\oplus}$ planets always exceed ${\gtrsim}6 \ R_{\oplus}$ and and lie at ${\sim}$0.4 days).

\subsection{Stellar Rotation as a Signature of Lossy RLO}\label{subsec:strot}

During top-down formation of a desert dweller, a significant amount of orbital angular momentum is transferred to the stellar spin through tidal interaction (equation \ref{equation:tides}). The stellar spin-up produced by ``lossy" RLO will also depend on the fraction of the lost angular momentum that is channeled into the star \citep[e.g.][]{pac81} as opposed to being expelled entirely from the system (which will not influence the stellar spin). Our first goal in this section is to ascertain whether the spin-up of desert dweller host stars may be observationally discernible. The second goal of this section is to determine whether stellar spin rates could encode information about where the angular momentum goes during lossy RLO (i.e. into the star or out of the system entirely). We will show that stellar spin-up may indeed be observable for ${\sim}10^{8-9}$ yr after the RLO episode. We will also demonstrate that stellar spins encode where the angular momentum is lost during mass transfer for a shorter time ${\lesssim}10^{8}$ yr after RLO.

We begin our analysis by constructing two flavors of stellar spin evolution tracks that bracket the possible degrees of spin-up. In the first set of tracks, we account only for tidal spin-up. In the second set, we additionally account for spin-up via accretion of material under the assumption that \textit{all} of the angular momentum that is not returned to the planet orbit is channeled into the host star. While in reality not all of the lost angular momentum may make its way into the star, we choose these two scenarios to showcase the least and most amount of spin-up possible. With this assumption, the stellar spin evolution contains an extra term (ignoring the ${\sim}{0.1\%}$ increase in $M_{\star}$ due to accretion of the ${\sim}300 \ M_{\oplus}$ of planetary mass lost):

\begin{equation}
    \dot{\omega}_{\star, \rm acc}=\frac{\Gamma|\dot{m}_{\rm RLO}|}{\alpha_{\star}R^{2}_{\star}}\sqrt{\frac{Ga}{M_{\star}}}.
\end{equation}

\begin{figure}
\epsscale{1.26}
\plotone{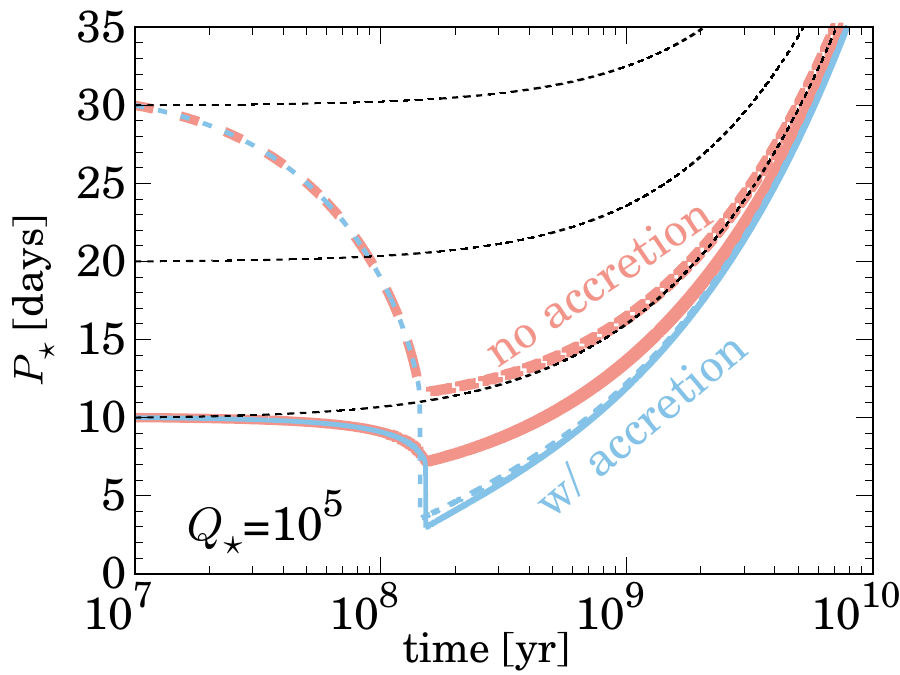}
\caption{Evolution tracks of the stellar rotation period $P_{\star}$. Rotation evolution for stars that only experience magnetic braking is shown in the three black dashed lines, corresponding to initial $P_{\star,0}{=}30,20,10$ days from top to bottom respectively \citep[metal-rich stars similar to those hosting desert dwellers exhibit a median rotation period ${\sim}22$ days;][]{amaroqmat20}. Stars that also experience tidal disruption of a giant planet with initial $P_{\star}{=}30$ days are shown in dashed curves, while those with initial $P_{\star}{=}10$ days in solid. Stars that only spin up due to tidal interaction are shown in red, while those that also spin up due to accretion of material are shown in blue. Desert dweller host stars could rotate faster than field stars of the same metallicity, mass, and age by as much as ${\sim}$an order of magnitude if observed shortly after RLO; even after ${\sim}$Gyr, rotation periods of desert dweller hosts may still differ from field stars by ${\sim}$10 days. \label{figure:st_spin}}
\end{figure}

We fix $Q_{\star}{=}10^{5}$ (varying $Q_{\star}$ shifts the timing of spin up but not the overall evolution) and choose two initial stellar rotation periods of $P_{\star,0}{\in}\{10,30\}$ days, roughly corresponding to slow and fast rotating main sequence FGK stars (\cite{mcqmazaig13}; see also Figure 6 of \cite{amaroqmat20} for the period distribution of main sequence metal-rich $0.85{-1} \ M_{\odot}$ stars similar to those of our desert dweller hosts). To study how long spin-up lasts, we also include a simple correction for magnetic braking whereby the stellar rotation velocity ${\propto}t^{-1/2}$ \citep[][]{sku72} via:

\begin{equation}\label{equation:skumanich}
    \dot{\omega}_{\star,\rm MB}=-\psi_{\rm MB} \omega^{3}_{\star},
\end{equation}

\noindent where $\psi_{\rm MB}{=}1.5{\times}10^{-14} \ \rm yr^{-1}$ \citep[e.g.][]{doblinmar04,valrapras15}. Equation \ref{equation:skumanich} may overestimate braking at high rotation rates similar to those achieved after mass transfer \citep{galbou15}. Our estimate for the timescale over which desert dweller host stars remain spun up is therefore a lower limit. For our runs with accretion, we integrate $\dot{\omega}_{\star}{=}\dot{\omega}_{\star,\rm tid}{+}\dot{\omega}_{\star,\rm acc}{+}\dot{\omega}_{\star,\rm MB}$.\footnote{We note that including magnetic braking or stellar accretion does not significantly change the results presented in Section \ref{sec:results}.}

Our spin evolution tracks are displayed in Figure \ref{figure:st_spin}. For slow rotators, tides dominate spin-up; tides alone are capable of transforming slow into rapid rotators with $P_{\star}{\sim}10$ days. Stellar accretion exacerbates tides, helping to shrink $P_{\star}$ by up to an order of magnitude for slow rotators. The difference between slow rotators that did experience RLO and those that did not (grey dashed lines in Figure \ref{figure:st_spin}) remains substantial ($\Delta P_{\star}{\sim}15{-}20$ days) even over ${\sim}$Gyr timescales.
Unsurprisingly, the imprint of RLO is much less distinguishable for initially rapid rotators. Similarly, the rotation periods of initially slow rotators that gained angular momentum from accretion and those that did not are significantly discrepant for ${\sim}\mathrm{a \  few}{\times}10^{8}$ yr after RLO, while for initially rapid rotators the difference is minor (a few days).

The upshot of Figure \ref{figure:st_spin} is that the spin periods of desert dweller host stars could encode an imprint from RLO only if they were not too rapidly rotating before RLO, i.e. exhibited periods $P_{\star}{\gtrsim}$20 days. Encouragingly, the stellar sample from \cite{amaroqmat20} for $0.85{-}1 \ M_{\odot}$ main sequence stars with [Fe/H]$\geq$0.1 indicates a typical $P_{\star}{\sim}22$ days (albeit with large scatter). This typical $P_{\star}$ value suggests that desert dweller host stars could rotate faster than field stars of the same metallicity, mass, and age by as much as ${\sim}$an order of magnitude if observed shortly after RLO. If observed ${\gtrsim}$1 Gyr after RLO, typical rotation periods may still differ by up to ${\sim}$10 days. Future work should address whether such period differences are indeed detectable given the large spread in field star rotation rates that could obscure the signal. RLO changes stellar spin evolution by roughly the same amount as variations in stellar metallicity \citep[][]{amamat20,amaroqmat20}, so spin-up should be equally detectable as the effect of metallicity with sufficient sample size. The Nancy Grace Roman Space Telescope's Galactic Bulge Time Domain Survey is projected to deliver ${\sim}10^{4-5}$ close-in, large transiting planets \citep[][]{wilbarpow23}, which could supply a sizable enough sample of desert dwellers to enable comparison of host vs. field star rotation periods.

\subsection{Direct Observational Signatures of Lossy RLO}

In this section we explore whether lossy RLO can be observed directly via luminosity generated during accretion/mass transfer. We first consider extra luminosity generated by an accretion disk comprised of material lost from the planet. If the disk is vertically optically thick (which we confirm to be true during RLO), it will radiate as a blackbody as it dissipates \citep[e.g.][]{lynpri74,frakinrai02}: 

\begin{align}\label{equation:lacc}
\begin{split}
    L_{\rm acc} &\sim \frac{3}{8\pi}\frac{GM_{\star}\dot{m}_{\rm acc}}{R_{\star}} \\
    & \sim 10^{-1} \bigg(\frac{\dot{m}_{\rm acc}}{10^{7} \ M_{\oplus} \ \rm Gyr^{-1}}\bigg) L_{\odot},
\end{split}
\end{align}

\noindent where we have evaluated for a solar-type star and used a value for $\dot{m}_{\rm acc}{\sim}|\dot{m}_{\rm RLO}|$ between the maxima from Figures \ref{figure:evolution_zoom} and \ref{figure:evolution_mc=20}. We assume $\dot{m}_{\rm acc}{\sim}|\dot{m}_{\rm RLO}|$ since the time for the disk to adjust to a new mass accretion rate, i.e. the viscous time $t_{\nu}{\sim}R^{2}/\nu{\sim}10 \ (R/R_{\odot})(\alpha_{\rm t}/10^{-2})^{-1}$ yr \cite[where we have used a disk alpha viscosity prescription $\nu{=}\alpha_{\rm t}c^{2}_{\rm s}/\Omega$;][]{shasun73}, is $\ll$ the timescale for changes in the mass transfer rate $t_{\dot{m}}{\sim}|\dot{m}_{\rm RLO}/\ddot{m}_{\rm RLO}|{\gtrsim}10^{4}$ yr.\footnote{The exception occurs for the planet in Figure \ref{figure:evolution_zoom}, which is the most extreme mass loss episode we consider. This case exhibits a minimum $t_{\dot{m}}{\sim}10$ yr which is comparable to the disk's viscous time for our fiducial turbulence level $\alpha_{\rm t}{\sim}10^{-2}$.} Given the disk effective temperature $T_{\rm e}{\sim}(L_{\rm acc}/R^{2}\sigma_{\rm SB})^{1/4}$, we find the emission should peak at wavelength $\lambda{=}hc/3k_{\rm B}T_{\rm e}{\sim}1 \mu m$, where $\sigma_{\rm SB}$ is the Stefan-Boltzmann constant, $c$ is the speed of light, $h$ is Planck's constant, and $k_{\rm B}$ is Boltzmann's constant. This estimate indicates that the emission should peak in the near infrared \citep[assuming it is not reprocessed to longer wavelengths by dust at larger distances from the star where it can condense, e.g.][]{adaladshu87}. 

\begin{figure}
\epsscale{1.26}
\plotone{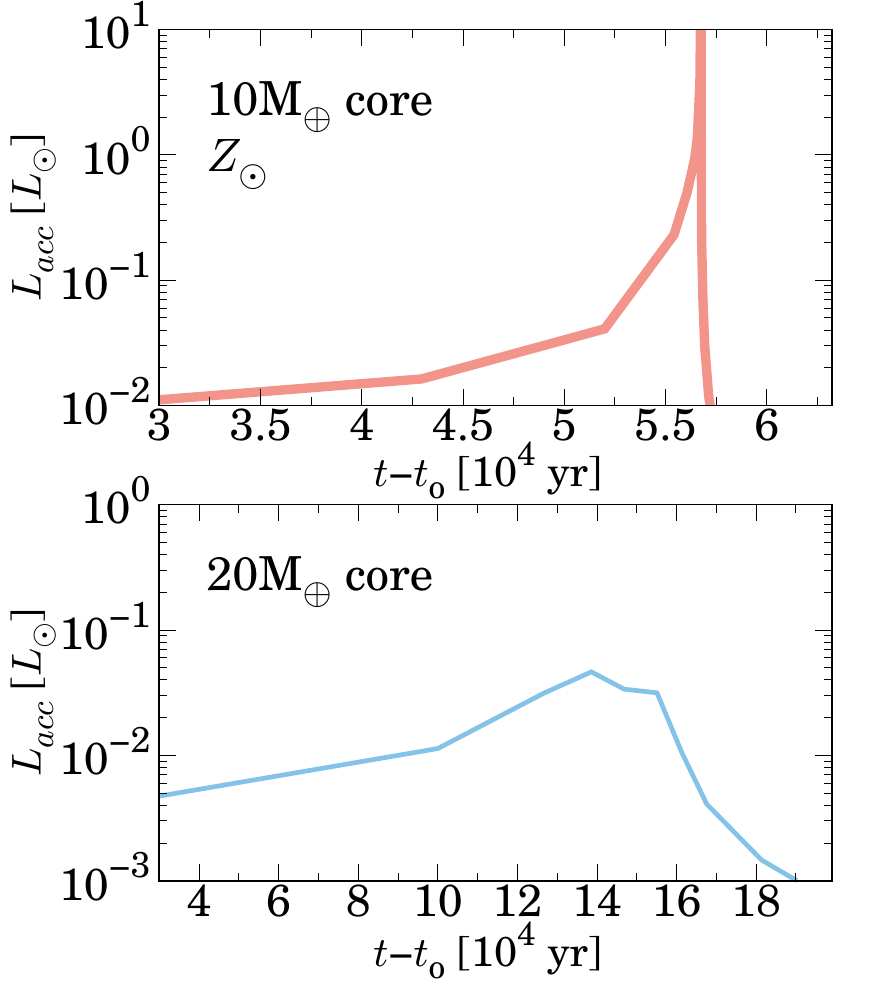}
\caption{Accretion luminosity (equation \ref{equation:lacc}) during lossy RLO for the planets shown in Figures \ref{figure:evolution_zoom} and \ref{figure:evolution_mc=20} on top and bottom panels, respectively. Lossy RLO may manifest as a burst of IR excess emission from the accretion disk; in the fiducial case we consider (i.e. upper panel; $10 \ M_{\oplus}$ core, $\Gamma{=}0.85$, $Z{=}Z_{\odot}$) the burst outshines the host star for ${\sim}10^{3}$ yr. Less intense episodes of RLO still produce excess luminosity that is ${\sim}1{-}10\%$ that of the host star for longer timescales ${\sim}10^{5}$ yr. Such transient events may be observable with current surveys. \label{figure:lacc}}
\end{figure}

Figure \ref{figure:lacc} illusrates that in the most extreme cases we consider with maximum $\dot{m}_{\rm RLO}{\gtrsim}10^{8} \ M_{\oplus} / \rm Gyr$, lossy RLO may therefore manifest as a burst of IR excess that could dominate the host star's emission. Such a burst should only last for ${\sim}10^{2-3}$ yr however \cite[if the star accretes, UV/optical emission from the stellar surface may accompany the IR; e.g.][]{kenhar87}. In less intense episodes of RLO (e.g. Figure \ref{figure:evolution_mc=20}), IR excess at a level ${\sim}1{-}10\%$ that of the stellar luminosity may persist for a longer time ${\sim}10^{5}$ yr. We note that an IR excess of ${\sim}1{-}10\%$ the stellar luminosity is comparable to that produced by protoplanetary disks commonly observed to accrete at ${\dot{m}}{\sim}10^{-8} \ M_{\odot} \ \mathrm{yr}^{-1}$ \citep[e.g.][]{rydzak87,kenhar87}. 

A similar burst of emission may be expected if the angular momentum loss during lossy RLO is brought about by an outflow from the surface of the star and/or disk. In this case, RLO may be accompanied by blueshifted metal forbidden lines and absorption features similar to T Tauri winds \citep[e.g.][]{haredwgha95,nisantalc18}. Thus, systems caught in the act of lossy RLO could present as particularly ``windy", despite their main sequence ages. Given the disruption rate of hot Jupiters in the Galaxy ${\sim}0.1{-}1 \ \rm yr^{-1}$ \citep[][]{metgiaspi12}, transient surveys may detect several such events over ${\sim}$decade-long baselines. Encouragingly, planetary tidal engulfment has already been serendipitously caught in the act by the Zwicky Transient Facility \citep{demackar23}. While engulfment events are more luminous than our RLO scenario, they are shorter lived ($<$ yr).  Current transient facilities thus likely already have the capacity to detect the onset of RLO in real time. 

Another strategy to observe RLO in real time is through targeted investigation of stars with anomalously depressed emission lines, since depression of common resonance lines (e.g. Ca II H $\&$ K) likely betrays the presence of ablating planetary material. Such a search strategy has indeed already found a handful of planets in the midst of mass loss \citep[in the Dispersed Matter Planet Project; ][]{hasstabar20}. Such a targeted search strategy may prove the most effective way to catch the death throes of hot Jupiters en route to becoming desert dwellers.

\subsection{Future Work}\label{subsec:future_work}

We close our discussion by itemizing what we regard as pressing open questions that can further elucidate the fate of hot Jupiters and the origins of desert dwellers.

Roughly ${\sim}{1/3}$ of our desert dwellers are confirmed to have external companions. These include TOI 4010 \citep[][]{kunvanhua23}, TOI 2000 b \citep[][]{shavanhua23}, TOI 1410 \citep[][]{pollubbea24}, TOI 1347 b \citep[][]{pollubbea24,rubdaihow24}, TOI 1288 b \citep[][]{knugannow23,pollubbea24}, WASP 84 c \citep[][]{macgollob23}, Kepler 94 \citep[][]{marisahow14,weiisahow24}, and Kepler 411 b \citep[][]{sunioagu19,weiisahow24}. A common feature of these planets with companions is that they lie near the lower and/or outer desert edge \citep[see also][]{doyarmacu25}. Future work should address whether the presence of companions in these systems is consistent or in tension with the picture developed here. At first blush, the presence of companions in these systems appears to be at odds with top-down formation, since most hot Jupiters appear to be lonely \citep[e.g.][]{huawutri16}. A sizable minority of hot Jupiters \textit{do} however boast companions \citep[at least $12{\pm}6\%$;][]{wuricwan23}; follow up statistical work would be helpful to determine whether the occurrence rate of desert dwellers with companions is consistent with that of hot Jupiters. Assessing whether these systems can maintain orbital stability and mutual observability through tidal destruction of a hot Jupiter may also serve as a powerful constraint on our theory \citep[see][for exploration of an inverse problem]{batbodlau16}. Alternative formation channels that do not involve hot Jupiters (e.g. bottom-up formation via major mergers) could potentially account for desert dwellers with nearby companions and should be investigated.

Hot Jupiters could possibly be disrupted by the high-eccentricity migration pathway as well as through tidal inspiral \citep[][]{liuguilin13}. Future theoretical work is needed to determine whether high eccentricity migration can indeed produce desert dwellers, and if so, how these two pathways can be differentiated through the properties of remnant planets they create (e.g. does high-eccentricity destruction produce desert dwellers with residual non-zero eccentricities?). Our current desert dweller sample contains only four planets with claimed eccentricity measurements, TOI 1288 b \citep[$e{=}0.064^{+0.014}_{-0.015};$][]{knugannow23} (though see \citep{pollubbea24} for a null detection), TOI 132 \citep[$e{=}0.059^{+0.05}_{-0.037}$;][]{diamatjen20}, TOI 908 \citep[$e{=}0.132{\pm}0.091$;][]{hawbayarm23}, and Kepler 411 b \citep[$e{=}0.146^{+0.005}_{-0.004}$;][]{sunioagu19}. Observational follow-up constraining the desert dweller eccentricity distribution would therefore also be helpful to distinguish between the two mechanisms. Differentiating between the tidal decay vs. high-eccentricity destruction scenarios could also shed light on the timing of hot Jupiter formation if the ages of desert dwellers can be well-constrained. 

Further work is also needed to build more realistic interior structure models, which we have shown play a central role dictating the outcome of mass transfer. Our models employ a core of fixed radius and a discontinuous boundary between core and gaseous envelope; in reality, the distinction between central ``core" and overlying ``atmosphere" is fuzzy \citep[][]{wahhubmil17,helste24}. Moreover, interiors may not be fully adiabatic \citep[e.g.][]{leccha13,manful21}. Understanding how compositional mixing and non-adiabaticity impacts planets' structural and thermal response to mass loss could be an important improvement on our calculations. Realistic treatment of these effects in one dimensional planetary evolution codes remains challenging however \citep[e.g.][]{mulhelcum20,sursutej24}.

Lastly, the fate of angular momentum removed from the planet orbit during RLO is currently unknown. A variety of loss mechanisms may play a role (see the discussion surrounding equation \ref{equation:adot_rlo}) and should be investigated. Determining where the angular momentum goes during mass transfer may elucidate the observational signatures of hot Jupiter destruction, and could inform whether tidal inspiral or alternative channels dominate desert dweller formation.

\section{Conclusion}\label{sec:conclusion}

Recent surveys have revealed a number of massive (${\sim}10{-}50 \ M_{\oplus}$) and dense ($\rho{\gtrsim}1 \ \rm g \ cm^{-3}$) planets residing in the previously uninhabited sub-Jovian desert \citep[e.g.][]{doyarmacu25}. Motivated by the fact that desert dwellers' host stars share the same metallicity distribution as that of hot Jupiters \citep{visbeh25}, in this work we have explored whether the desert could be populated by the remnants of tidally destroyed giant planets. We reiterate below our main conclusions:

\begin{itemize}
    \item Desert dweller host stars exhibit kinematics consistent with an older population descended from those hosting hot Jupiters.
    \item In order to produce desert dwellers of similar density to those observed, RLO must be ``lossy"; most of the angular momentum removed from the orbit by the outflow cannot be returned to the planet orbit. RLO is stable up to very high degrees of loss.
    \item Desert dwellers robustly emerge from ``lossy" RLO across a range of stellar tidal quality factors, planetary metallicities, and internal entropies. Initial entropy and the strength of tides are partially degenerate; some desert dwellers could emerge through weak stellar tides and a low initial entropy, or vice versa.
    \item The entire width of the sub-Jovian desert out to ${\sim}$3 days can be backfilled with the remnants of hot Jupiters that possessed their empirically inferred spread in entropies.
\end{itemize}

Our theory makes the following observational predictions that may be used to test the top-down formation picture:

\begin{itemize}
    \item The ultrahot Neptunes LTT 9779 b and TOI 3261 b should presently be undergoing tidal decay at a rate $|\dot{P}|{\sim}$0.5 ms/yr (assuming a constant $Q_{\star}$). 
    \item Desert dweller host stars may rotate up to an order of magnitude more rapidly than field stars with the same metallicity, mass, and age. Even ${\sim}$Gyr after RLO, rotation periods may differ by up to ${\sim}$10 days.
    \item Stars hosting planets undergoing lossy RLO may appear anomalously luminous or ``windy". Lossy RLO may manifest as IR excesses up to ${\sim}1{-}10 \ L_{\odot}$ for durations ${\sim}10^{2-3}$ yr, while lower excesses ${\sim}0.01{-}0.1 \ L_{\odot}$ may persist for longer periods ${\sim}10^{5}$ yr. Angular momentum loss via outflows from the star/disk may be distinguished via blueshifted features similar to T Tauri winds.
\end{itemize}

If our theoretical predictions are confirmed by observations, our calculations suggest that planets in the desert can be leveraged to study the interiors of giant planets in unprecedented detail.

\appendix{}

\section{Modified Bernoulli Integral}\label{sec:appendix_bernoulli}

In this section we outline how the calculated mass loss rates differ if variations in the mean molecular weight of outflowing gas are accounted for. We find that accounting for mean molecular weight differences does not significantly affect the mass loss rate.

We assume an isothermal wind of temperature $T$. The following Bernoulli integral is conserved along a streamline:

\begin{equation}\label{equation:bernoulli_full}
    \frac{1}{2}v^{2}+\int^{P}_{P_{\rm ph}} \frac{dP}{\rho}+\phi=\mathrm{const},
\end{equation}

\noindent where $v$ is the speed of outflowing gas and the integral extends from the planet photosphere. Assuming the gas is launched strongly subsonically so that $v(P_{\rm ph}){\ll}c_{\rm s}$, and setting $\rho{=}P{/} c^{2}_{\rm s}{=}P\mu(P)m_{\rm H}/(k_{\rm B}T)$, equation \ref{equation:bernoulli_full} implies that,

\begin{equation}\label{equation:root}
    \phi_{\rm L1}=\phi_{\rm ph}-\frac{1}{2}c^{2}_{\rm s, L1}-\int^{P_{\rm L1}}_{P_{\rm ph}}\frac{k_{\rm B}TdP}{\mu(P)m_{\rm H}P}.
\end{equation}

\vspace{1cm}
\noindent The mean molecular weight $\mu(P)$ varies with pressure according to H$_{2}$ dissociation equilibrium. We self-consistently calculate $\mu(P)$ from chemical equilibrium using the method outlined in Paper I. To find $\rho_{\rm  L1}$ and therefore compute the mass loss rate, we numerically solve for the $P_{\rm L1}$ at which equation \ref{equation:root} is satisfied using \texttt{scipy}'s \texttt{brentq} root-finding routine. We then convert pressure to density via $\rho_{\rm L1}{=}P_{\rm L1}/c^{2}_{\rm s,L1}$ and re-evaluate equation \ref{equation:mdot_rlo} to find $\dot{m}_{\rm RLO}$. We find that mass loss rates differ from the case without H$_{2}$ dissociation by order unity factors. Since the root-finding procedure outlined in this section is computationally slow, we ignore mean molecular weight variations in our fiducial calculations.

\section{Additional Parameter Study Plot}\label{sec:appendix_Q}

\begin{figure*}
\centering
\includegraphics[width=1\textwidth]{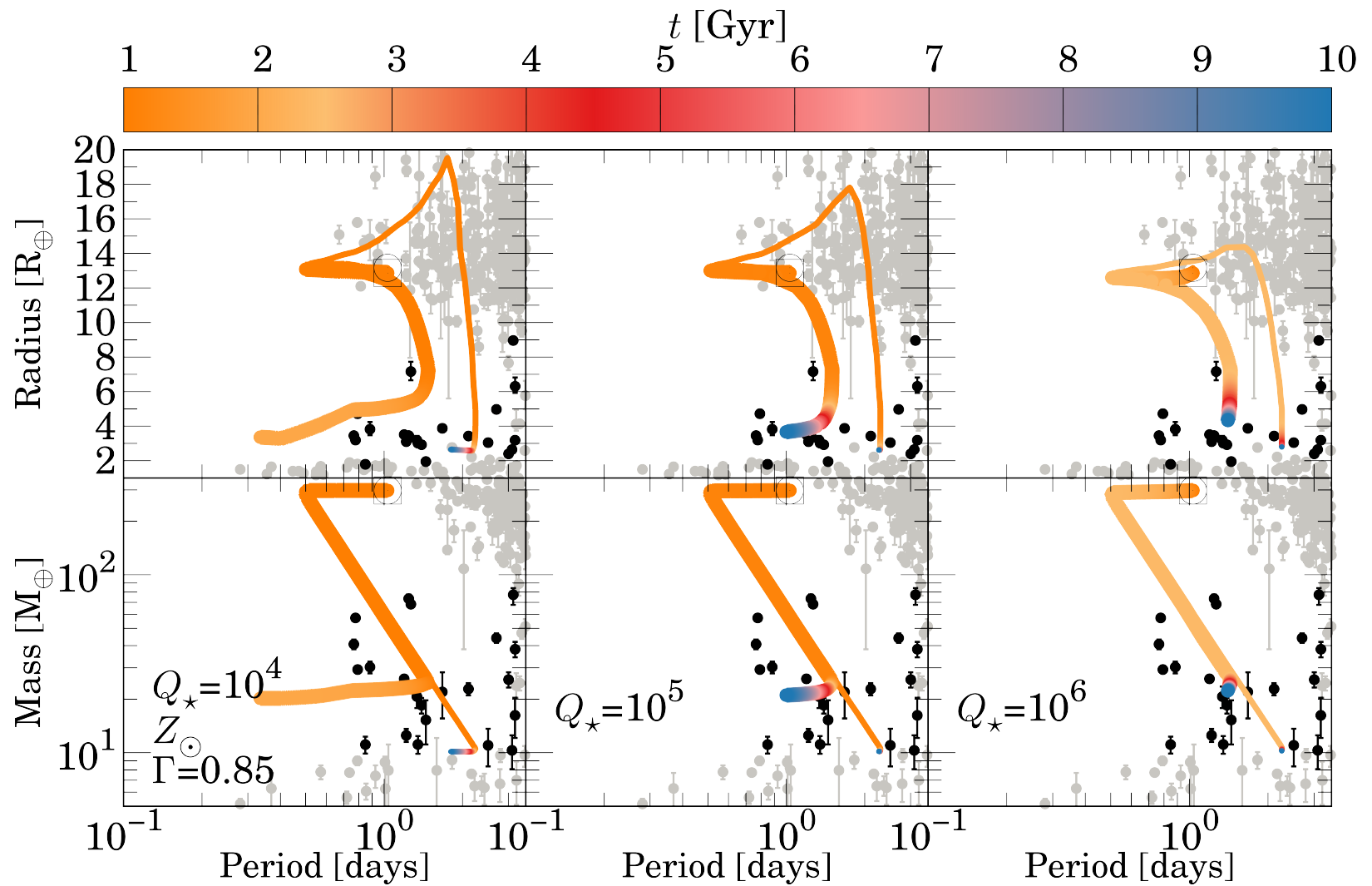}
\caption{Similar to Figure \ref{figure:rlo_3} except we vary the strength of stellar tides and keep planet metallicity and the degree of angular momentum loss constant. Evolution under stellar tidal quality factors $Q_{\star}{=}10^{4,5,6}$ is shown from left to right panels, respectively. The strength of tides does not significantly alter the RLO evolution, but does change the long-term evolution due to orbital shrinkage.
\label{figure:rlo_3_Qcomp}}
\end{figure*}

In Figure \ref{figure:rlo_3_Qcomp} we display additional evolution tracks isolating the effect of stellar tidal quality factor ($\log Q_{\star}{\in}[4,5,6]$). We find that the RLO evolution is largely unchanged, except for the long-term tidal decay of planets after mass transfer.

\acknowledgments

We thank the referee for a useful report that improved the paper. T.H. thanks David Armstrong, Andrew Cumming, Brian Jackson, Tommi Koskinen, Dong Lai, Doug Lin, Morgan Macleod, Christoph Mordasini, James Owen, Margaret Pan, James Rogers, Yao Tang, Shreyas Vissapragada, and Sam Yee for insightful and enlivening conversations. The authors acknowledge the MIT Office of Research Computing and Data for providing high performance computing resources that have contributed to the research results reported in this paper. Figures were created with \texttt{gnuplot}. Numerical integrations were carried out with \texttt{scipy} \citep[][]{virgomoli20}. 

\bibliography{hallatt}{}
\bibliographystyle{aasjournal}

\end{document}